\documentclass[twocolumn,showpacs,preprintnumbers]{revtex4}

\usepackage{graphicx}
\usepackage{dcolumn}
\usepackage{bm}
\usepackage{amssymb}
\usepackage{amsmath}
\usepackage{epsfig}

\def\th13 {\theta_{13}}

\def\bc {\begin{center}}
\def\ec {\end{center}}
\def\be {\begin{equation}}
\def\bea {\begin{eqnarray}}
\def\ee {\end{equation}}
\def\eea {\end{eqnarray}}

\def\lapp{\mathrel{\rlap{\raise.5ex\hbox{$<$}}
                    {\lower.5ex\hbox{$\sim$}}}}
\def\gapp{\mathrel{\rlap{\raise.5ex\hbox{$>$}}
                    {\lower.5ex\hbox{$\sim$}}}}

\def\dam{{(\Delta m_{31}^2)^m}}
\def\da{{(\Delta m_{31}^2)}}

\begin{document}
\title{Sensitivity to neutrino mixing 
parameters with atmospheric neutrinos}

\author 
{Abhijit Samanta
\footnote{E-mail address: abhijit@hri.res.in}}
\affiliation{Harish-Chandra Research Institute,
Chhatnag Road,
Jhusi,
Allahabad 211 019,
India}
\date{\today}

\begin{abstract}
\noindent
We have analyzed the atmospheric neutrino data  to study  the octant of $\theta_{23}$
and the precision of the oscillation parameters for a large Iron CALorimeter (ICAL) 
detector. The ICAL  being a tracking detector has the ability to measure the energy and 
the direction of the muon with high resolution. From  bending of the track in magnetic 
field it can also distinguish  its charge. We have generated  events  by  Nuance and then 
considered only the muons (directly measurable quantities) produced in charge current interactions
in our analysis. This encounters the main problem of
wide resolutions of energy and baseline. The energy-angle correlated 
two dimensional resolution functions are used to migrate the energy and the zenith angle of the neutrino
to those of the muon.
A new type of binning has been introduced to get better reflection of the oscillation pattern 
in chi-square analysis.
Then the marginalization of the $\chi^2$ over all parameters 
has been carried out for neutrinos and anti-neutrinos separately. 
We find that
the measurement of $\theta_{13}$  is possible at a significant precision
with atmospheric neutrinos.
The precisions of $\Delta m_{32}^2$ and $\sin^2\theta_{23}$ are found 
$\sim$ 8\% and 38\%, respectively, at  90\% CL. 
The discrimination of the octant as well as the deviation from maximal mixing of 
atmospheric neutrinos are also possible for some combinations of ($\theta_{23},
~\theta_{13}$). 
We also discuss the impact of the events at near horizon on the  precision studies.
\end{abstract}



\maketitle
\section{Introduction}
Recent discovery of the neutrino mass has opened up a new window into physics beyond 
the standard model. Aside from this fact, the two surprising sets of results \cite{Fogli:2008ig},
i) the extremely small masses of neutrinos 
(very different from quark sector)
and ii) a dramatically different mixing pattern 
from quarks, 
indicate  a new direction of this field. The first one may be the hint of
a new symmetry such as $B-L$ at high scales so that one can use a mechanism like seesaw 
to resolve the puzzle of the smallness of the masses. On the other hand, the second one
poses a much more challenging problem. One can expect a new symmetry for leptons as well as 
for quarks to solve this problem. Currently, there are many theoretical ideas. For example, the
 $\mu-\tau$
symmetry \cite{Mohapatra:2005yu} is invoked to explain the maximal mixing.  However, if this 
$\mu-\tau$ symmetry exists, it leads to $\delta_{CP}=0$ and $\theta_{13}=0$.  It should be 
broken if there appears a nonzero $\theta_{13}$ and CP violation. In that case the octant
(the sign of $\theta_{23}-45^\circ$) and the nonzero value of $\theta_{13}$ emerges other
new possibilities. It is also expected that neutrino theories may have implications
on the very fascinating fields like observed matter-antimatter asymmetry of the universe,  
grand unification, supersymmetry, extra dimensions, etc\cite{Mohapatra:2005wg}. 
 
Active endeavors  are under way to launch the era of precision experiments 
with a thrust
to uncover the underlying principle that gives neutrino masses 
and their mixing. This is one of the most promising  ways to explore physics
beyond the standard model.
In the standard oscillation picture there are six parameters.  
The present 1$\sigma$, 2$\sigma$ and 3$\sigma$ confidence level (CL) ranges
from global $3\nu$ oscillation analysis (2008)
\cite{Fogli:2008ig} are very exciting
(see table \ref{t:global-fit}).
Recently,  new bounds,  $\theta_{13}=-0.07^{+0.18}_{-0.11}$ and 
 the asymmetry $\theta_{23}-\pi/4=0.03^{+0.09}_{-0.15}$ at 90\% CL have
been shown in \cite{Escamilla:2008vq,Roa:2009wp} from an analysis  considering all
present neutrino data.
The CP-violating phase $\delta_{CP}$ is still unconstrained.

\begin{table*}[htb]
\begin{tabular}{cccccc}
\hline
Parameter & $\Delta m_{21}^2/10^{-5}\mathrm{\ eV}^2$ & $\sin^2\theta_{12}$ & $\sin^2\theta_{13}$ & $\sin^2\theta_{23}$ &
$|\Delta m_{31}^2|/10^{-3}\mathrm{\ eV}^2$ \\[4pt]
\hline
Best fit        &     7.67     &  0.312          &  0.016          &  0.466          &  2.39 \\
$1\sigma$ range & 7.48~--~7.83 & 0.294~--~0.331  & 0.006~--~0.026  & 0.408~--~0.539  & 2.31~--~2.50 \\
$2\sigma$ range & 7.31~--~8.01 & 0.278~--~0.352  & $<0.036$        & 0.366~--~0.602  & 2.19~--~2.66 \\
$3\sigma$ range & 7.14~--~8.19 & 0.263~--~0.375  & $<0.046$        & 0.331~--~0.644  & 2.06~--~2.81 \\
\hline
\end{tabular}
\caption{\sf \small Global 3$\nu$ oscillation analysis (2008)}
\label{t:global-fit}
\end{table*}

Despite of these spectacular achievements, a lot of things  are  
still missing. Tremendous efforts  are underway to determine the mass ordering
(sign of $\Delta m_{32}^2$),
the values of $\theta_{13}$ and $\delta_{CP}$, and to discriminate the octant degeneracy of 
$\theta_{23}$ 
in future experiments.
We define {$\Delta m^2_{32}=m_3^2-m_2^2$}.
There are many ongoing and planned experiments:  
UNO \cite{Jung:1999jq}, T2K \cite{Itow:2001ee}, NOvA \cite{Ayres:2004js}, 
 Hyper-Kamiokande \cite{Nakamura:2003hk},
INO \cite{Arumugam:2005nt}
and many others.
%
The main characteristic feature of a magnetized Iron CALorimeter (ICAL) detector 
proposed at India-based Neutrino Observatory (INO)  
is that it has the capability to detect $\nu_\mu$ and $\bar\nu_\mu$ separately, which
measures directly the matter effect.  


Unlike a fixed baseline neutrino beam experiment, the atmospheric neutrino flux covers a wide 
range of baseline (a few km -- 12900 km) and energy (sub GeV -- a few hundred GeV). 
On the other hand, it is not known well and there are huge uncertainties in its estimation.
It is also a  very rapidly falling function of energy. So, the extraction of the results from 
the experimental data is very complicated. 

The deviation from maximal mixing and the discrimination of octant degeneracy of $\theta_{23}$
have been studied in \cite{Choubey:2005zy,Indumathi:2006gr} with atmospheric neutrinos for a
large magnetized ICAL detector. 
However, the results have been obtained without marginalization and assuming the
Gaussian resolution functions of fixed widths for whole range of energy and zenith angle. 
The energy range for the atmospheric neutrinos
is very wide. The resolutions are changed significantly over its  range and are very different for 
neutrinos and anti-neutrinos. Moreover, the energy resolutions
appear to be non-Gaussian due to some unmeasurable product particles like neutral hadrons in neutrino 
interaction even if one considers all visible hadrons. 

For a given neutrino energy and direction, there is a distribution in the
reconstructed energy and direction. Again, a particular reconstructed energy
and direction can come from a wide range of true neutrino energy and direction.
So, it is not possible to convert a distribution in reconstructed energy and
direction obtained from an experiment  to a distribution in actual neutrino
energy and direction. This restricts the binning of the data for chi-square analysis
only in experimentally measured energy and direction.
On the other hand, the actual resolution functions have no regular pattern
and significantly deviate
from the Gaussian nature even if we consider the visible hadrons.
Again, the width changes with neutrino energy. For a simplistic
analysis, if  one considers a Gaussian resolution with a width that gives equal
space under the surface of resolution function, the
correct theoretical data smearing this approximated Gaussian resolution function
can not be obtained
for chi-square analysis. As a consequence, the best-fits and the contours of oscillation parameters
will  differ largely from the true values. In literature, there are many analyses where
both the theoretical as well as the experimental data are obtained by smearing
the Gaussian resolution functions. For an example, see ref. \cite{Indumathi:2006gr}. However, the result
changes very rapidly with change of the width of the resolution.
So, realistic estimation of the capability of an experiment can be done only by
an analysis with experimentally measurable quantities
and exact resolution functions.


Till now, the precision studies with atmospheric neutrinos have mainly carried out  for water
Cherenkov detector, a non-magnetized detector.
It is very important to see the capability of a large magnetized detector. 
We have studied the neutrino oscillation
considering neutrinos and  anti-neutrinos separately in the chi-square analysis.
Here, we consider the muons (directly measurable quantities at ICAL) produced by the charge current 
interactions. We generate  events by Nuance-v3 \cite{Casper:2002sd}. 
The two dimensional
 energy-angle correlated resolution functions are used to migrate the energy and the zenith angle
of the neutrino to the energy and the zenith angle of the muon. 

The above method has been introduced in \cite{Samanta:2006sj} and later used in
 \cite{Samanta:2008ag}. 
The goal of the previous work \cite{Samanta:2008ag}  was only to compare the allowed
parameter space of oscillation parameters obtained from different types of binning.
The considered systematic uncertainties were very much different from the present systematic
uncertainties. The purpose of this work is to study the following.

We consider whole data set in previous studies. But in reality, the horizontal events cannot 
be detected when the iron slabs are stacked horizontally.
In this paper, we have studied the impact of these events  
 in determining the precision of the parameters with and without considering  a rejection criteria 
for the horizontal events. This is very crucial to determine whether horizontal stacking of iron plates
is better than the vertical stacking or not.
 
As discussed in  \cite{Samanta:2008ag}, the binning of the data neither in 
$\log E-\cos\theta_{\rm zenith}$ nor in $\log E-\log L$ is the optimum. 
In this paper, we have optimized the binning in $L$. These are equal binned grids in 
$\log E - L^{0.4}$ plane, which can capture the oscillation behavior for all $L$ and $E$ 
in a better way in the chi-square analysis. 
Again, the number of bins in both axes need optimization between resolutions and statistics.
However, it should be noted here that if the statistics is huge for whole range of $E$ and $L$,
one can solve this problem by making the bin size very small and then the type of binning will
not play any crucial role. However, the type of binning is very crucial when the analyses is 
in experimentally measurable energy and directions. Here the statistics over measured energy and direction
is redistributed notably from  the true neutrino energy and direction.

Finally, we have made a detailed study on the sensitivity of a magnetized  
ICAL detector in determining the precision of $\Delta m_{32}^2$ and $\theta_{13}$ as well as in 
discriminating the octant ambiguity of $\theta_{23}$. 
We find the sensitivities of the parameters in two dimensional parameter space  after marginalization 
over whole allowed ranges of the parameters. The absolute bounds of each parameter are also studied.



\section{Atmospheric neutrino flux and events}\label{s:flux}
The atmospheric neutrinos are produced by the interactions of the
cosmic rays mainly with nucleuses of molecules in the earth's atmosphere. 
The knowledge of primary spectrum 
of the cosmic rays has been  improved from the observations by 
BESS\cite{Maeno:2000qx} and AMS\cite{Alcaraz:2000ss}. However,  large regions 
of parameter space have not been explored and they are interpolated 
or extrapolated from the measured flux.  The difficulties and the uncertainties
in the calculation of the neutrino flux depend on the neutrino energy.
The low energy fluxes have been known quite well. The 
cosmic ray fluxes ($<$ 10 GeV)  are modulated by the solar activity
and the geomagnetic field through a rigidity (momentum/charge) cutoff.
At the higher neutrino energy ($>$ 100 GeV), the solar activity and the rigidity 
cutoff are irrelevant\cite{Honda:2004yz}.  
There is 10\% agreement among the calculations for neutrino energy below 10 GeV
because different hadronic interaction models are used in the
calculations and because 
the uncertainty in the cosmic ray flux measurement
is 5\% for the cosmic ray energy below 100 GeV \cite{Honda:2004yz}. 
In our simulation, we have used a typical Honda flux calculated in 
3-dimensional scheme\cite{Honda:2004yz}. 

The  interactions of neutrinos with the detector material are simulated 
using the Monte Carlo model Nuance (version-3)\cite{Casper:2002sd}. 
Here, the charged current (CC) and 
neutral current (NC) interactions are considered for (quasi-)elastic, 
resonance, coherent, diffractive, and deep inelastic scattering  processes.

\begin{figure*}[htb]
\includegraphics[width=19.0cm,height=14cm,angle=0]{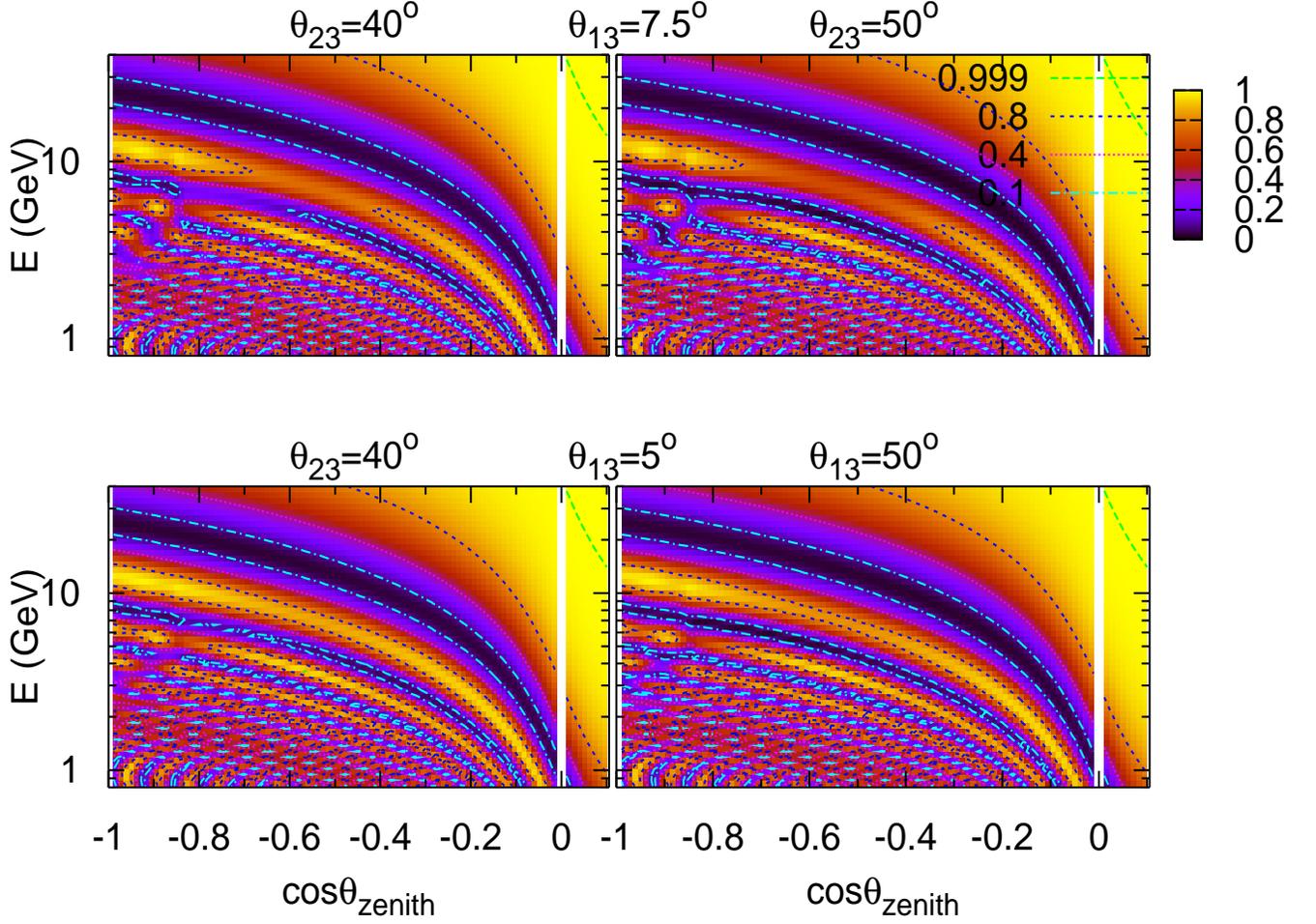}
\caption{\sf \small The oscillogram  of $\bar\nu_\mu\rightarrow\bar\nu_\mu$
oscillation probability in $E-\cos\theta_{\rm zenith}$ plane for $\theta_{23}=40^\circ$ (left column)
and $50^\circ$ (right column) with $\theta_{13}=5^\circ$ (lower row) and $7.5^\circ$ (upper row).
We choose $\Delta m_{32}^2=-2.5\times 10^{-3}$eV$^2$ and $\delta_{CP}=0$.}
\label{f:octant}
\end{figure*}

\section{Oscillation of atmospheric neutrinos}

The present atmospheric neutrino data are well explained by two flavor 
oscillation \cite{Ashie:2005ik, Ashie:2004mr}.
However, one expects a considerable $\nu_\mu\rightarrow\nu_e$ 
oscillation of atmospheric neutrinos in 3-flavor framework if $\theta_{13}$ is nonzero.  
To understand the analytical solution one may adopt the 
so called ``one mass scale dominance" (OMSD) frame work: 
$|\Delta m_{21}^2| << | m_{3}^2 -m_{1,2}^2|$.
Then the oscillation probabilities can be expressed as:
\begin{eqnarray}
\mbox{P}_{\mu e} &=& \mbox{P}_{e\mu} \nonumber \\
&=& \sin^2 \theta_{23} \sin^2 2 \theta_{13}
      \sin^2 \left(\frac{1.27 \Delta m_{31}^2 L}{E}\right); \nonumber \\
\mbox{P}_{\mu\mu} &=& 1 \nonumber \\
&& - 4 \cos^2 \theta_{13}
\sin^2 \theta_{23} ( 1-\cos^2 \theta_{13} \sin^2 \theta_{23}) \nonumber \\
&&\times 
\sin^2 \left(\frac{1.27 \Delta m_{31}^2 L}{E}\right).
\label{eqn:oscillation-vacuum}
\end{eqnarray}
These oscillation probabilities are derived for vacuum.  Since the oscillation 
involves electron neutrino, it will be modulated by the matter
effect \cite{Mikheev:1986gs,Wolfenstein:1977ue}. Then,
\begin{eqnarray}
\mbox{P}^m_{\mu e} &=& \mbox{P}^m_{e\mu} \nonumber \\
&=& \sin^2 \theta_{23} \sin^2 2 \theta_{13}^m
      \sin^2 \left(\frac{1.27 \Delta (m_{31}^2)^m L}{E}\right). \nonumber \\
\label{e:oscillation-matter}
\end{eqnarray}
Here,  $E$, $L$ and $\Delta m^2_{31}$ are in GeV, km and  eV$^2$, respectively.


\bea
P^{m}_{\mu \mu} &=&
{1 - \cos^2 \theta^m_{13} \; {\sin^2 2 \theta_{23}}} 
\nonumber\\
&&
\times\sin^2\left[1.27 \;\left(\frac{\da + A + \dam}{2}\right) \;\frac{L}{E} \right]
\nonumber \\
&& ~-~
\sin^2 \theta^m_{13}\; \sin^2 2 \theta_{23}
\nonumber \\
&& 
\times\sin^2\left[1.27 \;\left(\frac{\da + A - \dam}{2}\right) \;\frac{L}{E}
\right]
\nonumber \\
&& ~-~
{{\sin^4 \theta_{23}}} \;
{ { \sin^2 2\theta^m_{13} \;
\sin^2 \left[1.27\; \dam  \;\frac{L}{E} \right].
}}
\label{e:pmumu}
\eea

 The  mass squared difference ${{\dam}}$ and mixing angle
${ {\sin^22\theta_{13}^m}}$ in matter are related to their vacuum values by
\bea
\dam =
\nonumber
\sqrt{(\da \cos 2 \theta_{13} - A)^2 +
(\da \sin 2 \theta_{13})^2},
\\
sin2\theta^m_{13}=
\frac{{\da \sin 2 \theta_{13}}}
{\sqrt{(\da \cos 2 \theta_{13} - A)^2 +(\da \sin 2 \theta_{13})^2,} }
\label{e:dm31}
\eea

where, $A = 2\sqrt{2}G_FN_e E $, 
 \(G_F\) is the Fermi constant, \(N_e\) is the electron density
of the medium and \(E\) is neutrino energy \cite{Giunti:1997fx}.
The matter potential term \(A\) has the same absolute value, but opposite
sign for neutrino and anti-neutrino.
The superscript `m' denotes effective parameters in matter.
Due to this matter effect, the Mikheyev-Smirnov-Wolfenstein (MSW) resonance
occurs in $\mbox{P}(\nu_\mu \rightarrow \nu_e)$ or  
$\mbox{P}(\nu_e \rightarrow \nu_\mu)$. 
It happens for Normal Hierarchy (NH)
with neutrinos and for Inverted Hierarchy (IH) with anti-neutrinos. 
It can be understood from Eq. \ref{e:pmumu} and \ref{e:dm31} that
a resonance in above oscillation probabilities will occur for 
neutrinos (anti-neutrinos) with NH (IH) 
when
\be \sin^22\theta_{13}^m \rightarrow 1
 ~~~~{\rm or,}~~ A= \Delta m^2_{31} \cos2\theta_{13}. \ee
Then the resonance energy can be expressed as 
\bea E = \left [\frac{1}{2\times 0.76 \times 10^{-4} Y_e} \right]
\left [\frac{| \Delta m_{31}^2|}{\rm eV^2} \cos2\theta_{13} \right ]
\left[\frac {\rm gm/cc}{\rho} \right ].
\label{e:resonance}\eea
The resonance energy corresponding to a baseline can be seen 
in  \cite{Samanta:2006sj}.

The oscillogram of muon survival probability is demonstrated in Fig. \ref{f:octant}
for $\theta_{13}=5^\circ$ and $7.5^\circ$ with $\theta_{23}=40^\circ$ and $50^\circ$, respectively.
Here, we show the resonance ranges for the neutrinos passing through the core of the
earth (with $E \approx 3-6$  GeV) and the mantle of the earth (with $E\approx 5-10$ GeV).
We also see a difference for $\theta_{23}=40^\circ$ and $50^\circ$ due to the 
$\sin^4\theta_{23}$ term (Eq. \ref{e:pmumu}), which dominates over the other terms due to the matter effect.

\begin{figure*}[htb]
\includegraphics[width=12.0cm,angle=270]{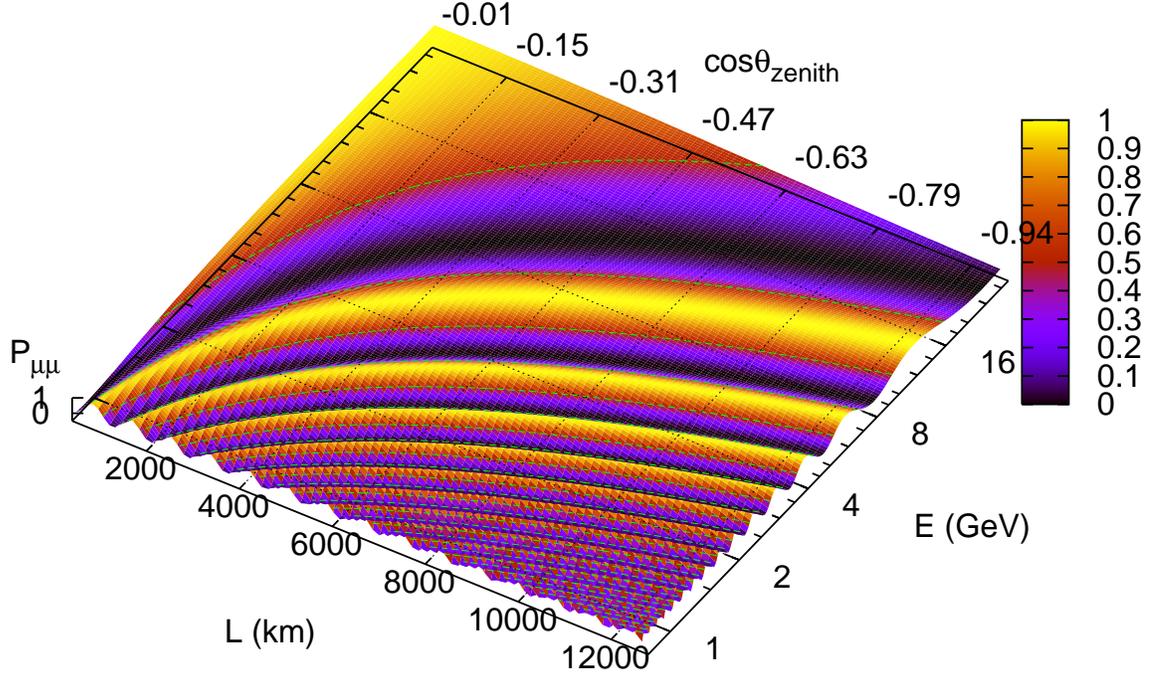}
\caption{\sf \small The  $\nu_\mu\rightarrow\nu_\mu$
oscillation probability in vacuum. We choose $\Delta m_{32}^2=-2.5\times 10^{-3}$eV$^2$,
$\theta_{23}=45^\circ$ and $\theta_{13}=0^\circ$.}
\label{f:p3dn}
\end{figure*}

\section{Binning of the events} 
For binning of the data in $E$, we need to consider the following facts.
I) The atmospheric neutrino flux falls very rapidly with increase in 
energy.
II) Again, the wide resolutions of $E$ and $L$ 
between true neutrinos and reconstructed neutrinos smear the 
oscillation effect to a significant extent. 
The wide resolutions arise mainly due to the interaction kinematics.
This huge uncertainty in reconstructed neutrino momentum is due to 
the un-observable product particles  and slightly 
due to the un-measurable momentum of recoiled nucleus when $E\lapp 1$ GeV. For this reason, the energy resolutions
deviate largely from the Gaussian nature. These are strongly 
neutrino energy dependent. 
At low energy ($E\lapp 1.5$ GeV) the quasi-elastic process
dominates and the muon carries almost whole energy of the neutrino. The energy resolution
is very good here.  With increase in energy, the 
width of the resolution  increases significantly as the deep 
inelastic event dominates as well as 
the flux also falls very rapidly.
This is one of the main problems in the atmospheric neutrino experiments.
III) There is also an important characteristics of the oscillation probability when 
both $L$ and $E$ are varied simultaneously.
We explain it here for $\nu_\mu \leftrightarrow 
\nu_\mu$ oscillation  in vacuum, which  is a sinusoidal function of $L/E$.
If we plot it in $L-E$ plane (see Fig. \ref{f:p3dn}),
it is seen that the distance between two consecutive 
peaks of oscillation in $E$ for a fixed $L$ increases very rapidly with $E$. 
These three points suggest  increase in bin size  with increase in $E$.
We choose equal bin size in $\log E$. 
 
Again, the distance between two consecutive peaks
of oscillation in $L$ for a fixed $E$ increases rapidly
as we go to lower values of $L$. When this distance is very small compared to the resolution width of
$L$, the oscillation effect is averaged out. Only when the distance is large, it contributes to
oscillation measurements. 
To get the reflection of this oscillation pattern 
in $\chi^2$,  we need decreasing bin size of $L$  with decreasing of its  value.
This has been studied in detail for three common choices 
of binning of the data in \cite{Samanta:2008ag}, and it has been found that neither 
$\log E-\cos\theta_{\rm zenith}$ nor in $\log E-\log L$ is optimum.
In this work we optimize the binning the data in the  grids  of
$\log E-L^{0.4}$ plane. 

The number of bins used for this analysis is discussed later in the section \ref{s:chi}.
Here, it should be noted  that one cannot
make the bin size arbitrarily small. The number of event in a bin may be a fraction
of 1 in theoretical data for chi-square analysis, but the number of event in  experimental data is 
either zero or integer number greater or equal to 1. Obviously, no chi-square method will work
if many of the bins have number of  event  equal to zero or just equal to 1. However, the 
number of events per bin $\ge 1$ is not also sufficient. We have 
checked that one needs number of  event per bin at least $>4$ to obtain $\chi^2$/d.o.f$\approx 1.$
This indicates the optimization of bin size with statistics.

\begin{figure*}[htb]
\bc
\includegraphics[width=5cm,angle=270]{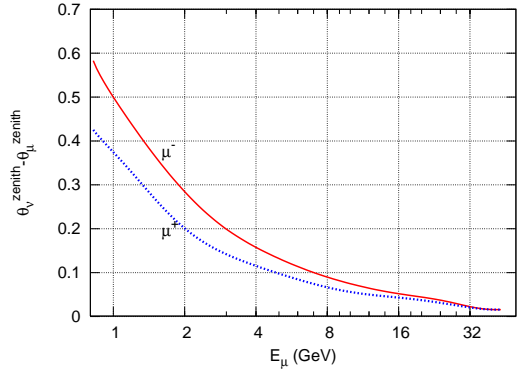}
\ec
\caption{\sf\small
The variation  of the half width at half maxima with $E_\mu$ for the 
distribution of ${\theta_\nu}^{\rm zenith}-{\theta_\mu}^{\rm zenith}$ at horizon.
The distribution is obtained for each  $E_\mu$ bin
from 500 years un-oscillated atmospheric data of 1 Mton ICAL.
}
\label{f:thrhalfmax}
\end{figure*}

\section{Selection of events}\label{s:cut}
The up going and down going events are mixed at the near horizon due the uncertainty in scattering
angle between  neutrino and muon. The up going neutrinos get oscillated and down going neutrinos
remain almost un-oscillated due to the short distance from the source to the detector. 
When the iron plates
of ICAL detector are placed horizontally, all these events cannot be detected. The high
energy events will have normally small scattering angle, but very long tracks in the detector.
So, they may be detected. If we plot the distribution of the difference in zenith angles 
between neutrino and the corresponding muon  for a fixed energy, it gives a 
Gaussian plot. The half width at the half maxima of this distribution as a function of 
muon energy is shown in fig \ref{f:thrhalfmax}. We put a selection criteria that the events 
for a given muon energy having the difference $|90^\circ-\theta_{\rm zenith}|$ within the above 
half width are rejected. Here we expect  roughly that these events cannot be detected
in case of real experiments.
The precisions determined with and without this cut is discussed later.

\section{The $\chi^2$ }\label{s:chi}
The number of events falls very rapidly with increase in energy and there is a very small 
statistics at the high energy. However,  the contributions to the sensitivities
of the oscillation parameters is significant from these high energy events.
For the low statistics at the high energy, the $\chi^2$ is calculated 
according to the Poisson probability distribution  defined by the  
expression:
 \begin{eqnarray}
 \chi^2 &=& \sum_{i,j=1}^{n_L,n_E} \left[ 2 \left\{ N^{p}_{ij} \left(
 1+\sum_{k=1}^{n_s} f^k_{ij} \cdot
 \xi^k \right) - N^{o}_{ij} \right\} \right. \nonumber \\
  && \left. - 2 N^{o}_{ij} \ln \left( \frac{N^{p}_{ij} \left( 1+
 \sum_{k=1}^{n_s} f^k_{ij} \cdot \xi_k \right) } {N^{o}_{ij}}\right)  \right] \nonumber \ \\
  && + \sum_{k=1}^{n_s} {\xi_k}^2
 \label{e:chisq}
 \end{eqnarray}

Here, $N^o_{ij}$ is  the number of observed events generated 
by Nuance 
for a given set of oscillation parameters with an exposure of 1 Mton.year of ICAL and
 $N^p_{ij}$ is the number of predicted events (discussed later). 
These are obtained in a 2-dimensional grids
in the plane of $\log E - L^{0.4}$.  The term $f^k_{ij}$ is the systematic uncertainty
of $N^p_{ij}$ due to the $k$th uncertainty (discussed later) and ${\xi_k}$ is the pull
variable for the $k$th systematic uncertainty. 
We use total number of $\log E$ bins $n_E$ = 35 (0.8 $-$ 40 GeV) and the number of  $L^{0.4}$ bins
as a function of the energy.
We consider $n_L = 2\times 25,~2\times 27,~2\times 29,~2\times 31,$ and $~2\times 33$ for
$E= 0.8-1,~ 1-2,~ 2-3,~ 3-4,~ {\rm and} ~ >4$ GeV, respectively.
For the down-going events, the  binning  is done
by replacing `$L^{0.4}$' by $`-L^{0.4}$'.
The factor `2' is taken to consider both up and down going cases.
For the up going neutrino, $L$ is the distance traveled by the neutrino
from the source at the atmosphere to the detector in the underground. In case of the down going 
neutrino, the $L$ is the `mirror $L$' which is 
the same $L$ if the neutrino comes from exactly opposite direction. 

The table of Honda flux is given in 20  $\cos\theta_{\rm zenith}$ bins and 101 $E$ bins 
($0.1 - 10^{4}$ GeV).
It should be noted that we first re-binned the data 
into 300  $\cos\theta_{\rm zenith}$ bins and 200 $\log E$ bins ($0.8-40$ GeV)  to get the 
oscillation pattern accurately. This large number of $\cos\theta_{\rm zenith}$ bins also help
in  proper re-binning of the data into $L^{0.4}$ bins. 

\subsection{Migration from neutrino to muon}
\begin{figure*}[htb]
\hspace*{-2cm}
\includegraphics[width=6.2cm,angle=270]{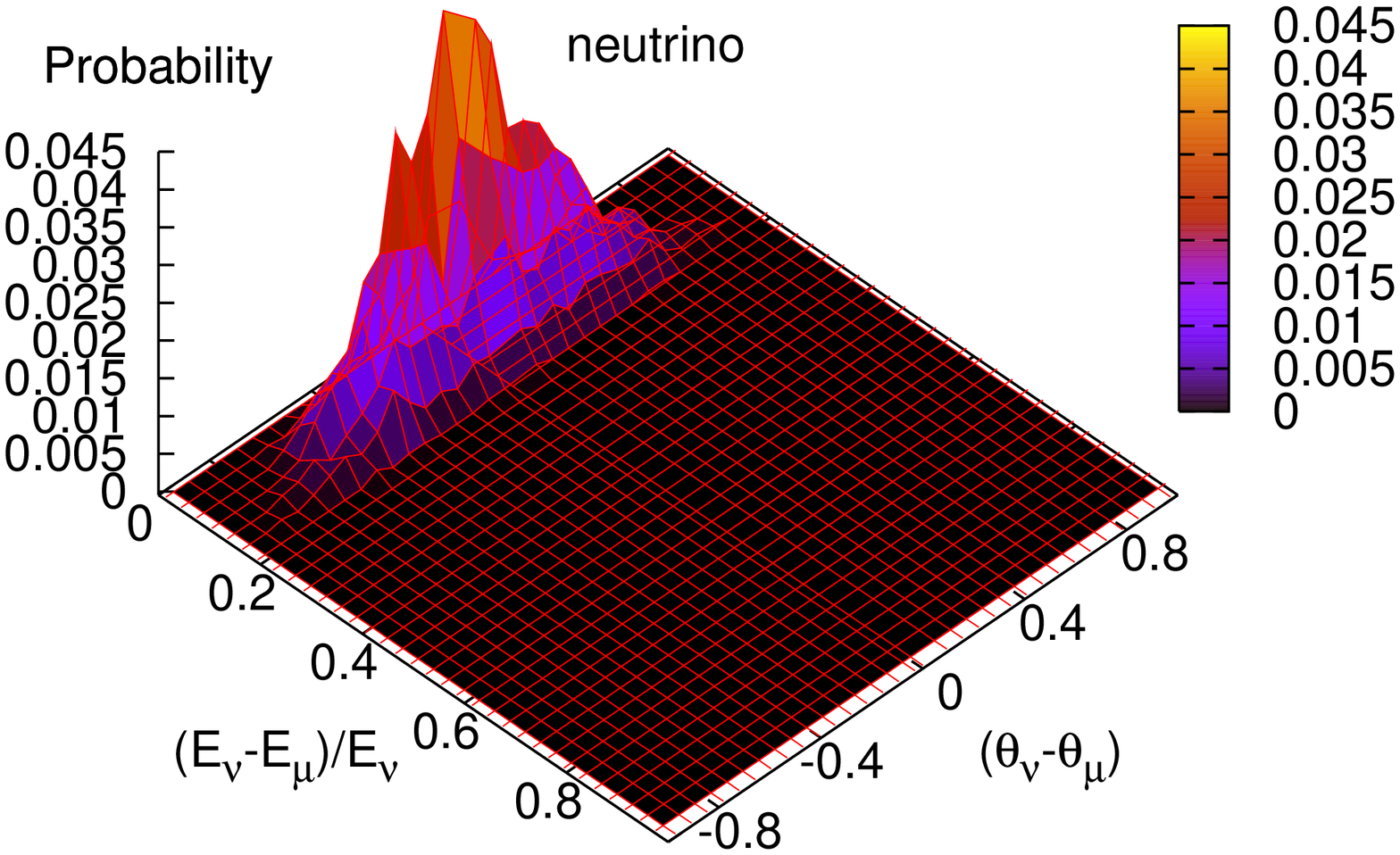}
\includegraphics[width=6.2cm,angle=270]{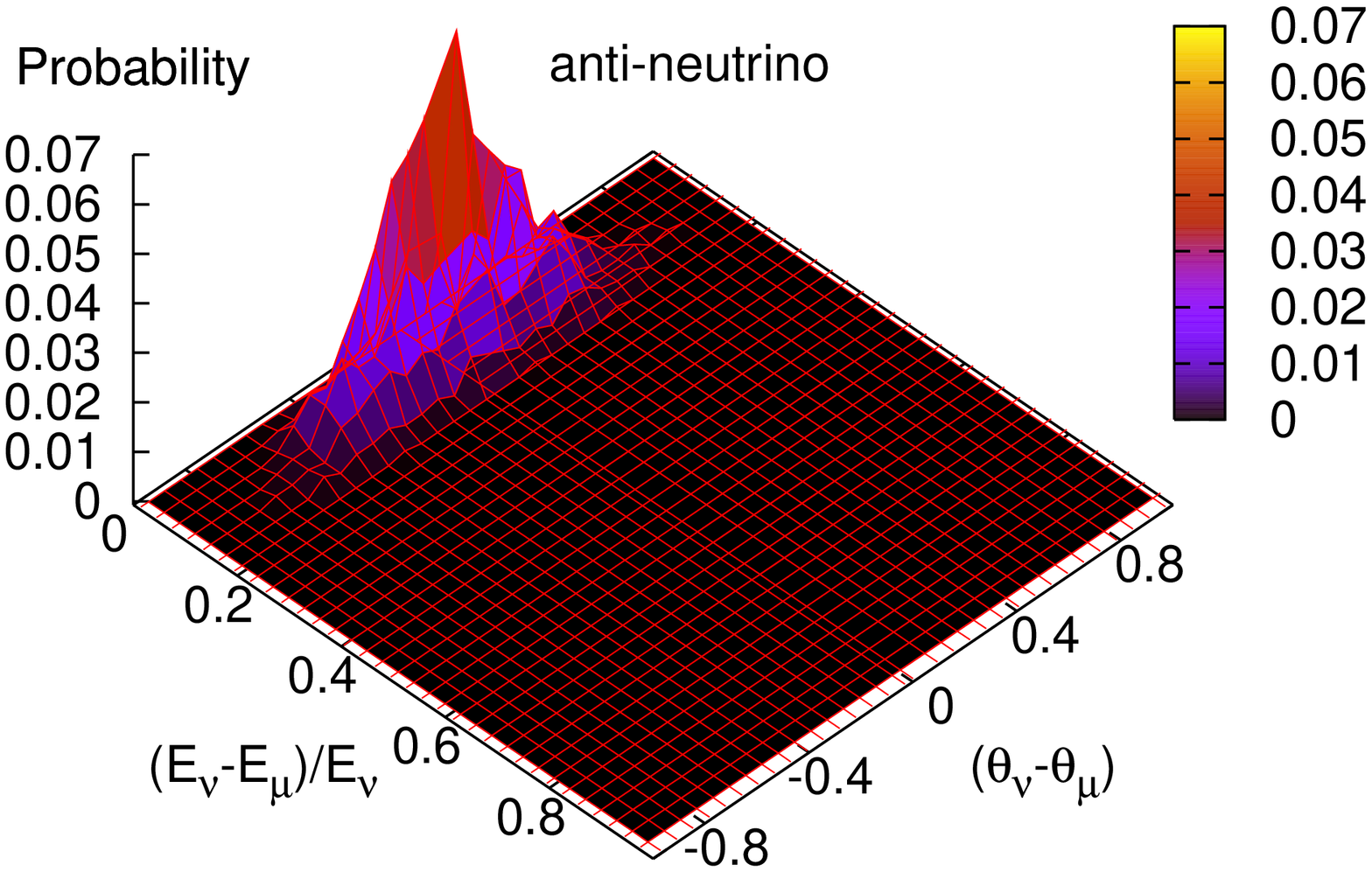}
\hspace*{-2cm}
\includegraphics[width=6.2cm,angle=270]{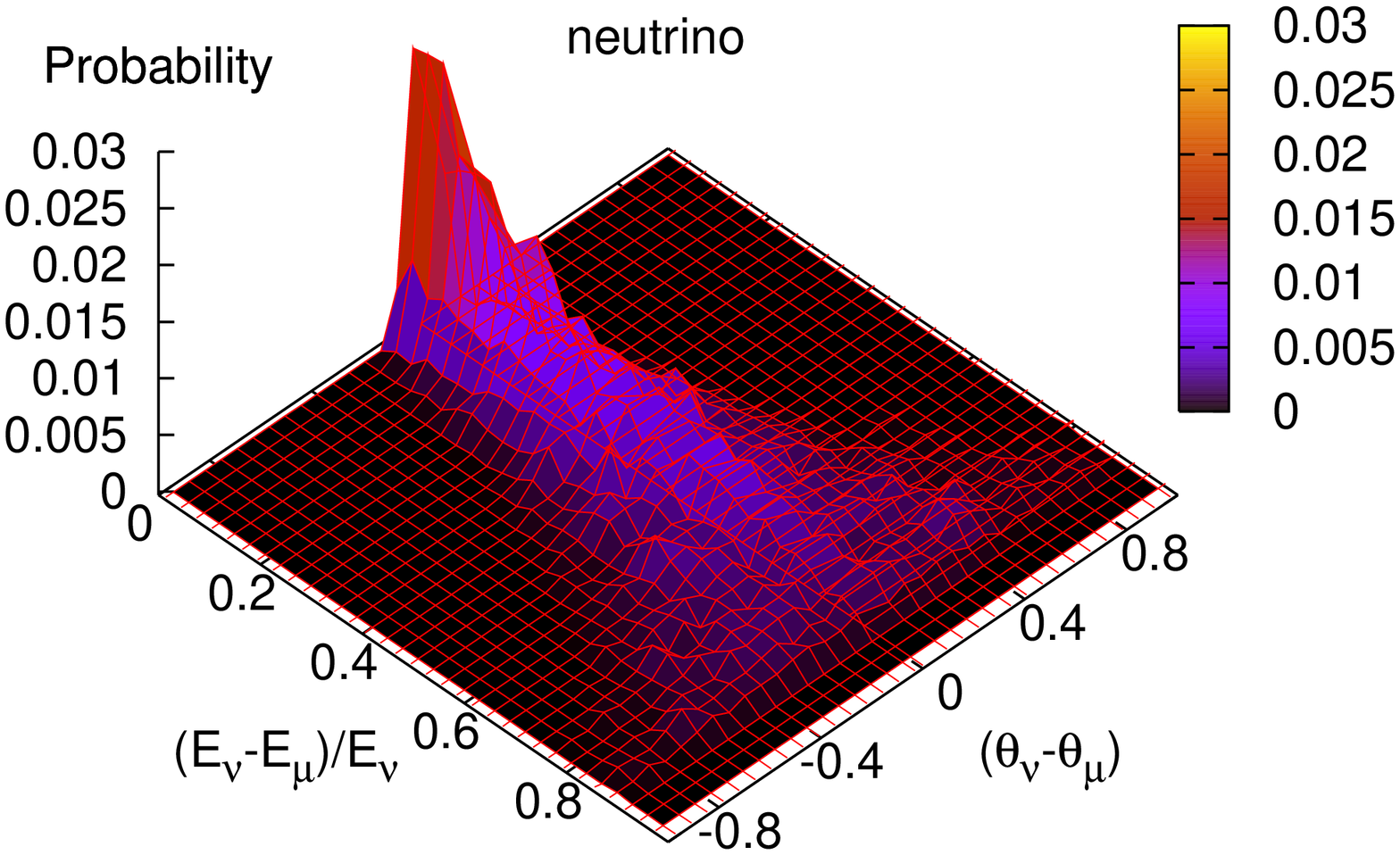}
\includegraphics[width=6.2cm,angle=270]{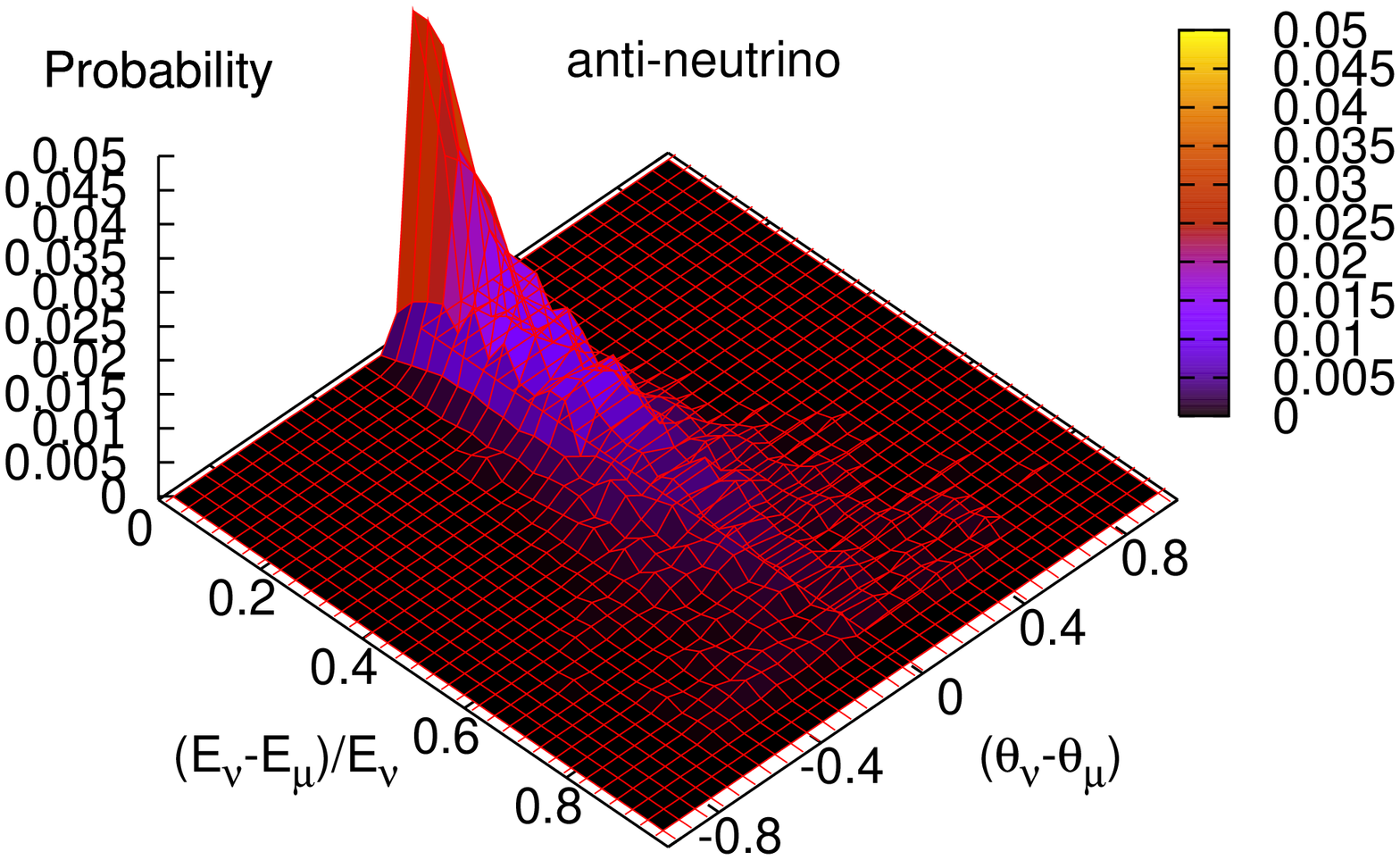}
\caption{\sf \small The sample energy-angle correlated resolution plots for neutrino 
(left column) and 
anti-neutrino (right column) for the bins of $E_\nu=0.85-0.98$ GeV with
$\cos\theta_{\rm zenith}= -0.40$ to  $-0.20$ (upper row) and 
$E_\nu= 6.84 - 7.86$ GeV with $\cos\theta_{\rm zenith}=0$ to $0.20$ (lower row).
The data is obtained from the simulation of 500 MTon.year
exposure of ICAL considering no oscillation.}
\label{f:reso}
\end{figure*}

To generate the theoretical data for the chi-square analysis,
we first generate 500 years un-oscillated data for 1 Mton detector by Nuance.
From this data we find the energy-angle correlated resolutions (see Figs. \ref{f:reso}) 
in 35 $E^\nu$ bins (in log scale for the range of $0.8-40$ GeV) 
and 17 $\cos\theta_{\rm zenith}^\nu$ bins (for the range $-1$ to $+1$). For a given $\log E_\nu$ bin,
we calculate the efficiency of having $E_\mu \ge$ 0.8 GeV (threshold
of the detector).
For each set of oscillation parameters, we integrate the oscillated 
atmospheric neutrino 
flux folding the cross section, the exposure time, the target mass, the efficiency 
and the resolution function to obtain the predicted data.
We use the CC cross section of Nuance-v3 \cite{Casper:2002sd} and the  
Honda flux  in 3-dimensional scheme \cite{Honda:2004yz}.
This method has been discussed in detail in \cite{Samanta:2006sj}, but the number of 
bins and resolution functions have been changed here. 

One can do this directly by generating 500 Mton.year data (to ensure
that the statistical error is negligible) for {\bf each set} of oscillation 
parameters and then reducing it to 1Mton.year equivalent data, which would 
be the more straight forward method. 
The marginalization study with this method is almost 
an un-doable job in a  normal CPU. However, an exactly equivalent result is 
obtained here using the energy-angle correlated resolution function.

We have done this study for ideal muon detector.  From GEANT simulation 
of ICAL detector it is seen that the energy resolution of muon varies 4--10\%
depending on the direction and energy. Since the iron plates are stacked 
horizontally, the resolution will be better for vertical events than the slanted
events. The angular resolution varies from 4--12\% for the considered range of
energy and  zenith angle.
Here, the thickness of iron plates are considered to be 6 cm.
From Fig. \ref{f:reso} it is clear that these are negligible compared to the
resolutions obtained from kinematics of scattering processes.  
 
The addition of the hadron energy to the muon energy of an event, which might improve
the reconstructed neutrino energy resolution, is not considered here for
conservative estimation of the sensitivity. It would be realistic in case
of GEANT-based studies since the number of hits produced by the
hadron shower  strongly depends on the thickness of iron layers.
However, ICAL can also detect the neutral current events. Though it is
expected that these events will not have any directional information; energy
dependency of the oscillation, averaged over all directions can also contribute
to the total $\chi^2$ in the sensitivity studies separately.



\subsection{Systematic uncertainties}
The atmospheric neutrino flux is not known precisely,  there are huge 
uncertainties in its estimation. We may divide them into two categories: 
I) overall uncertainties (which are independent of 
energy and zenith angle), and
II) tilt  uncertainties (which are  dependent of  energy and/or zenith angle).
We consider the following types of uncertainties.

The energy dependent uncertainty, which arises due to the uncertainty in 
spectral indices, can be expressed as 
\begin{equation}
      \Phi_{\delta_E}(E) = \Phi_0(E) \left( \frac{E}{E_0} \right)^{\delta_E}
      \approx \Phi_0(E) \left[ 1 + \delta_E \log_{10} \frac{E}{E_0} \right].
\label{e:uncer}
\end{equation}

Similarly, the vertical/horizontal flux uncertainty as a function of  zenith angle  can be expressed as
\begin{equation}
      \Phi_{\delta_z}(\cos\theta_z)
      \approx \Phi_0(\cos\theta_z) \left[ 1 + \delta_z (|\cos\theta_z|-0.5) \right].
\end{equation}
Next, we consider the overall flux normalization uncertainty $\delta_{f_N}$,
and the overall neutrino cross section uncertainty $\delta_{\sigma}$.

For $E < 1$ GeV we consider $\delta_E=5\%$ and $E_0=1$ GeV 
and for $E>10$  GeV, $\delta_E=5\%$ and $E_0=10$ GeV.
We take $\delta_{f_N}=10\%$, $\delta_{\sigma}=15\%$.
We consider $\delta_z=4\%$ which leads to 2\% vertical/horizontal flux uncertainty.  We derived these uncertainties from \cite{Honda:2006qj}. 

For each set of oscillation parameters, we calculate the $\chi^2$ in two stages.
First we used $\xi_k$ such that $\frac{\delta \chi^2}{\delta \xi_k}=0$, which 
can be obtained solving the  equations \cite{Fogli:2002pt}. Then we calculate the final $\chi^2$ with these $\xi_k$ values.
Finally, we minimize the $\chi^2$ with respect to all oscillation parameters
\footnote{Here, we consider all uncertainties as a function of reconstructed 
neutrino energy and zenith angle. Here we assumed that the tilt uncertainties 
will not be changed too much due to reconstruction. However, on the other 
hand if any tilt uncertainties 
arises in reconstructed neutrino events from the reconstruction method or the
kinematics of the scattering, it is then accommodated in $\chi^2$.}.

\section{Marginalization and Results}
A global scan of $\chi^2$ is carried out over the oscillation parameters 
$\Delta m_{32}^2,~\theta_{23}$, $\theta_{13}$ and $\delta_{CP}$  
with neutrinos and anti-neutrinos separately.
We have chosen the range of $|\Delta m_{32}^2|=2.0-3.0\times 10^{-3}$eV$^2$,
$\theta_{23}=38^\circ - 52^\circ$, $\theta_{13}=0^\circ - 12.5^\circ$ and 
$\delta_{CP}=0^\circ - 360^\circ$.
The 2-dimensional 68\%, 90\%, 99\% 
confidence level allowed parameter spaces (APSs) are obtained by considering 
$\chi^2=\chi^2_{\rm min}+2.48,~4.83,~9.43$, respectively. 
For every set of data we have checked that chi-square/d.o.f remains $\lapp 1.1$ at its minimum value.
We obtain the APS in $|\Delta m_{32}^2| - \theta_{23}$ and $|\Delta m_{32}^2| - \theta_{13}$ plane. 
We  set the input of $|\Delta m_{32}^2|=2.5\times 10^{-3}$eV$^2$ and $\delta_{CP}=0$.

It is important to note here that the statistics changes significantly over $L-E$ plane with the 
change  of oscillation parameters. 
Moreover, the fluxes and the resolutions are very different at different $L-E$ zones.
The upper and lower bounds of an oscillation parameter
depends significantly on the statistics as well as on the resolutions of the specific zones 
in $L-E$ plane. The binning of the data, which captures the oscillation patten
also plays the vital role. 

However, for some sets of input parameters the chi-square remains 
almost flat over a significant range of a parameter and then changes rapidly. 
It happens due to the fact that 
I) the  change of oscillation probability   is insignificant, and/or
II) the above change is significant, but it is eaten by the 
systematic uncertainties in chi-square analysis.
In this circumstances, the best-fit values may change significantly from the input values.
This is a very common feature in analyses with generating events by Monte Carlo 
method. But, in methods without Monte Carlo,  the number of events are determined with an accuracy
of a fraction of 1 and then best-fit values is always close to the input values. 

\begin{figure*}[htb]
\includegraphics[width=10cm,angle=0]{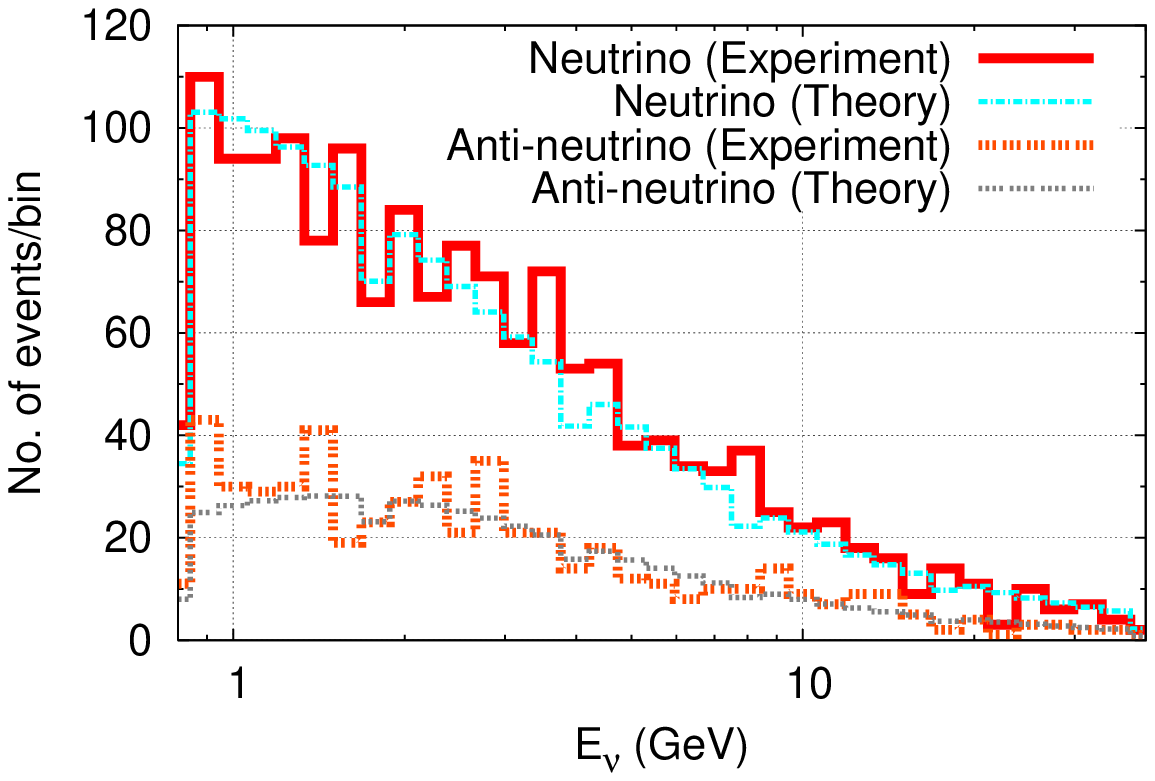}
\includegraphics[width=10cm,angle=0]{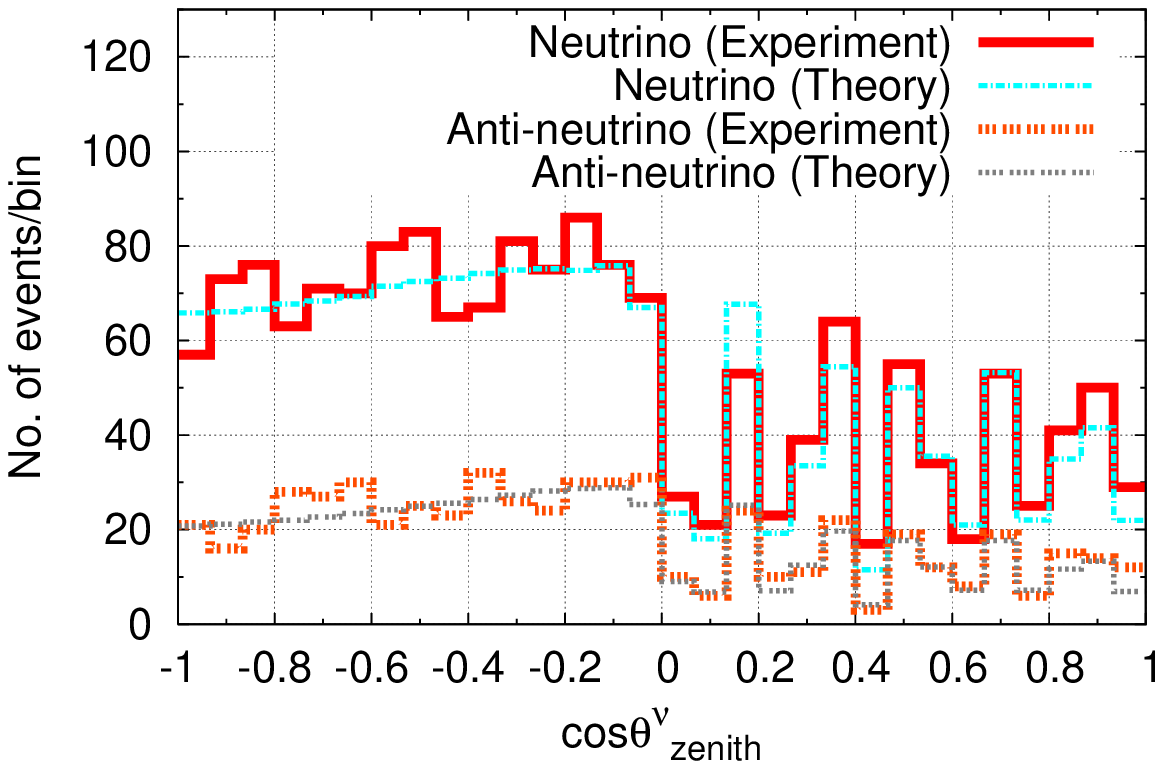}
\caption{\sf \small
The typical distribution of events with $E_\nu$ keeping $\cos\theta^\nu_{\rm zenith}$ fixed 
at $\approx -0.367$  and with $\cos\theta^\nu_{\rm zenith}$ keeping $E_\nu$ fixed at $\approx 2.24$ GeV.
We set $\Delta m^2_{32}=-2.5\times 10^{-3}$eV$^2$, $\theta_{23}=45^\circ$, $\theta_{13}=0^\circ$
and $\delta_{CP}=0^\circ$. 
}
\label{f:event}
\end{figure*}

In some cases, the deviations of the best-fit values are large. This is  happened
due to the following reasons.
Here, we have just folded the total charge current cross section of all processes
to find the number events for a particular neutrino energy to generate the theoretical
data. We see significant
fluctuations more than 1 $\sigma$ in  number of events between ``theoretical data"
and ``experimental data" in some particular energy bins for a given set of oscillation
parameters (see Fig. \ref{f:event}). This happens mainly at the neutrino energy $\lapp 3$ GeV,
where the resonances occur. Here, the neutrino cross sections depend on the type of nucleus.
The generation of events is very complicated here and it depends on the models.
These all are not considered in the same way as in Nuance in generation of
theoretical data, which causes  energy dependent systematic uncertainty.
However, this has no regular pattern. In our analysis we consider only the over
all uncertainty in the cross section.
These energy dependent uncertainties have not been considered in our analysis.
When $\theta_{23}$ deviates from $\pi/4$, the difference between peak and dip decreases
and the fluctuations
becomes relatively prominent. Again, when $\theta_{13}$ becomes large, the periodic pattern of
oscillation is lost due to matter effect.
We have checked that the fluctuations are larger
for $\theta_{23}=50^\circ$ and $\theta_{13}=7.5^\circ$
than $\theta_{23}=45^\circ$ and $\theta_{13}=0^\circ$.
In this region of oscillation parameters,  significant deviations
of best-fit values of oscillation parameters from their true values are obtained.

\begin{figure*}[htb]
\includegraphics[width=6.0cm,angle=0]{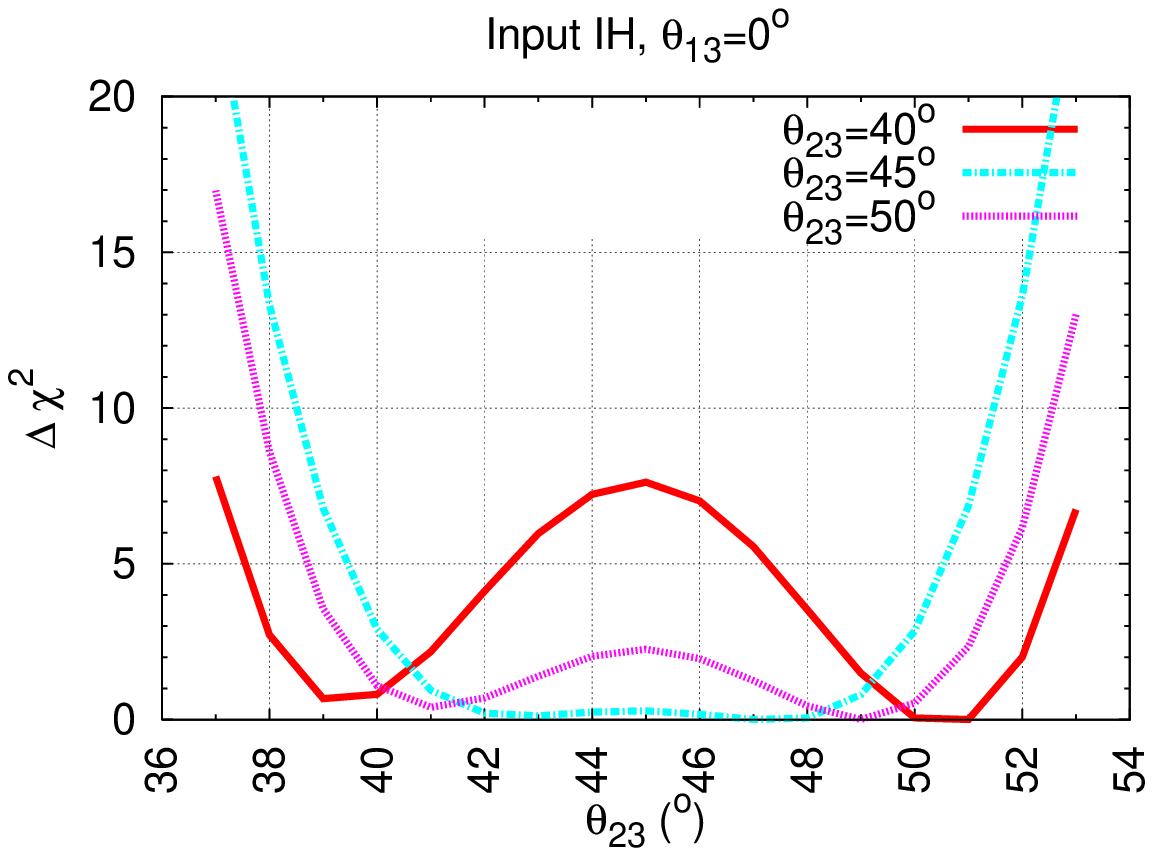}
\includegraphics[width=6.0cm,angle=0]{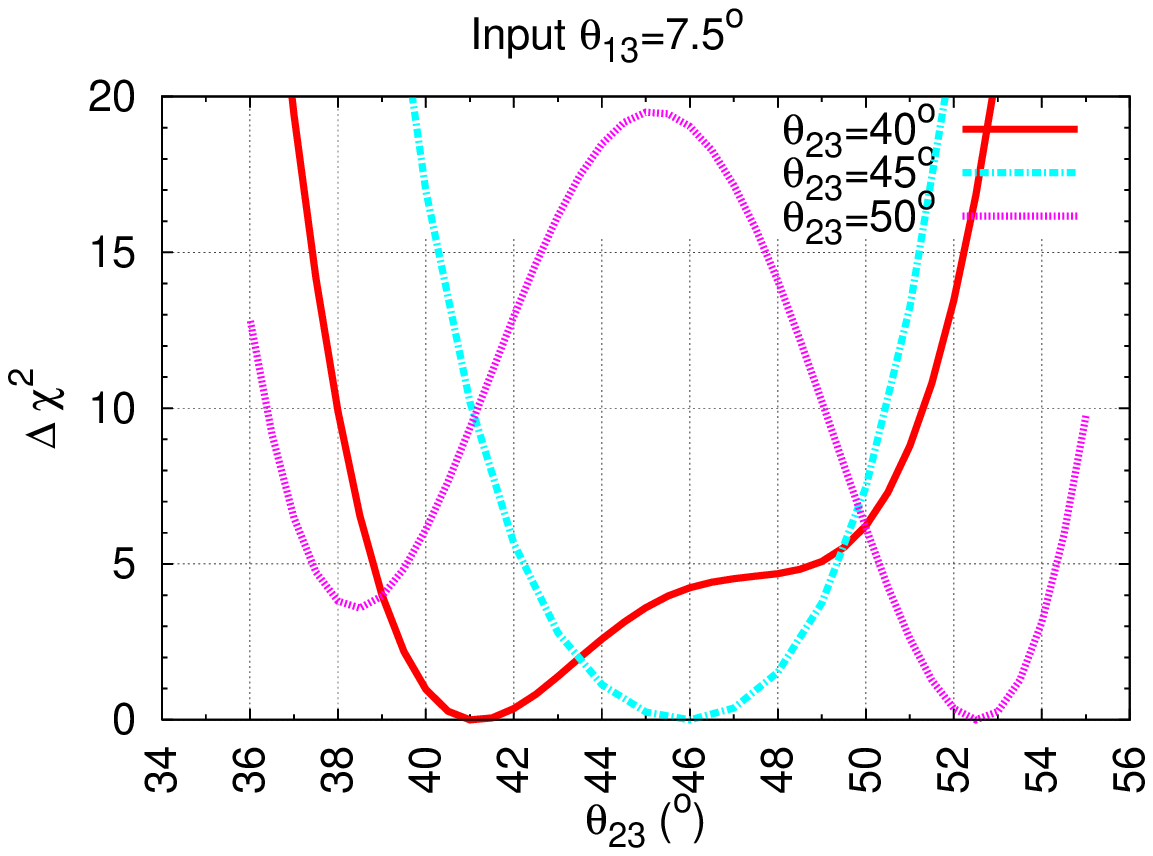}
\caption{\sf \small 
The variation  of $\Delta \chi^2 = (\chi^2-\chi^2_{\rm min})$ with $\theta_{23}$
for input value of $\theta_{23}=40^\circ$, $45^\circ$ and $50^\circ$ with
$\theta_{13}=0^\circ$ and $\theta_{13}=7.5^\circ$, respectively. Here, we consider
both neutrinos and anti-neutrinos together. The type of input hierarchy is inverted.
}
\label{f:tt23}
\end{figure*}
\begin{figure*}[htb]
\includegraphics[width=6.0cm,angle=0]{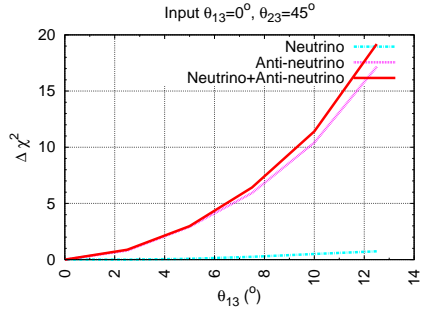}
\caption{\sf \small
The variation  of $\Delta \chi^2 = (\chi^2-\chi^2_{\rm min})$ with $\theta_{13}$
for input value of $\theta_{13}=0^\circ$, 
and $\theta_{23}=45^\circ$ considering neutrino, anti-neutrino and both types of neutrinos, respectively.
}
\label{f:tt13ns}
\end{figure*}
\begin{figure*}[htb]
\includegraphics[width=6.0cm,angle=0]{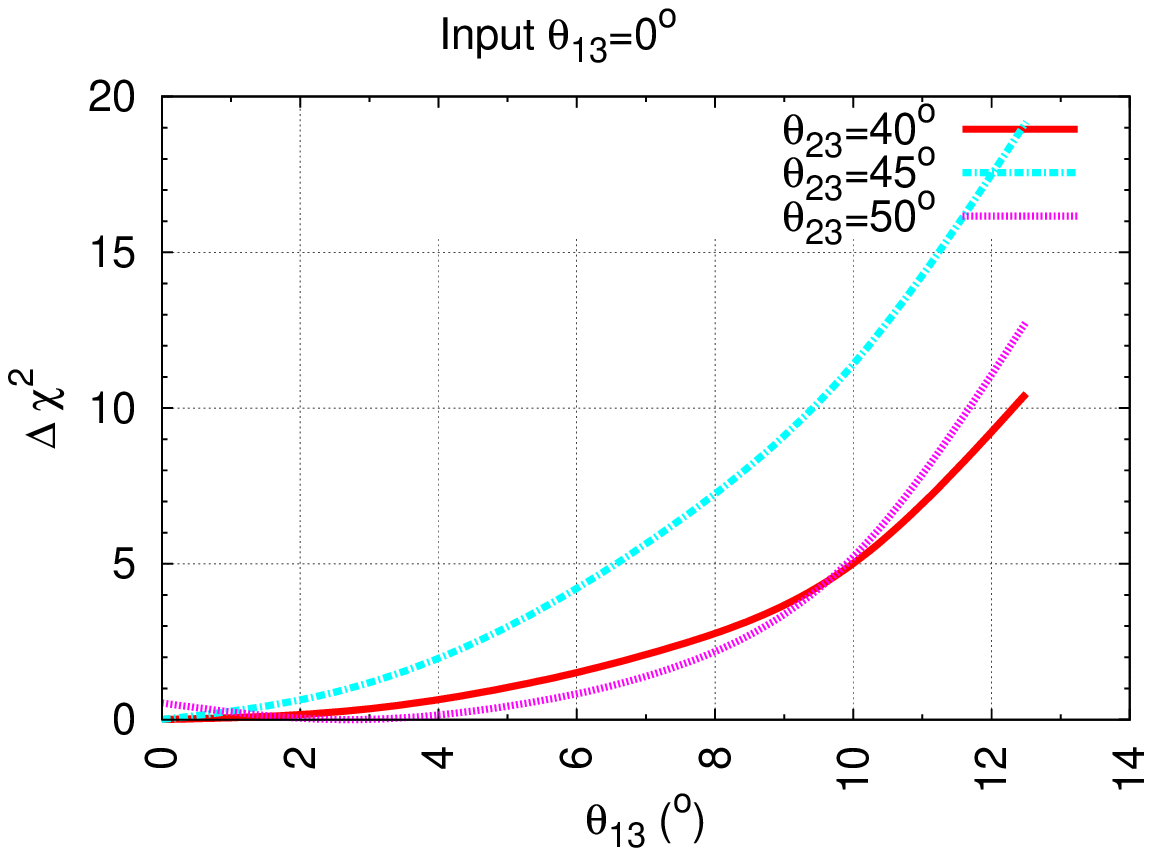}
\includegraphics[width=6.0cm,angle=0]{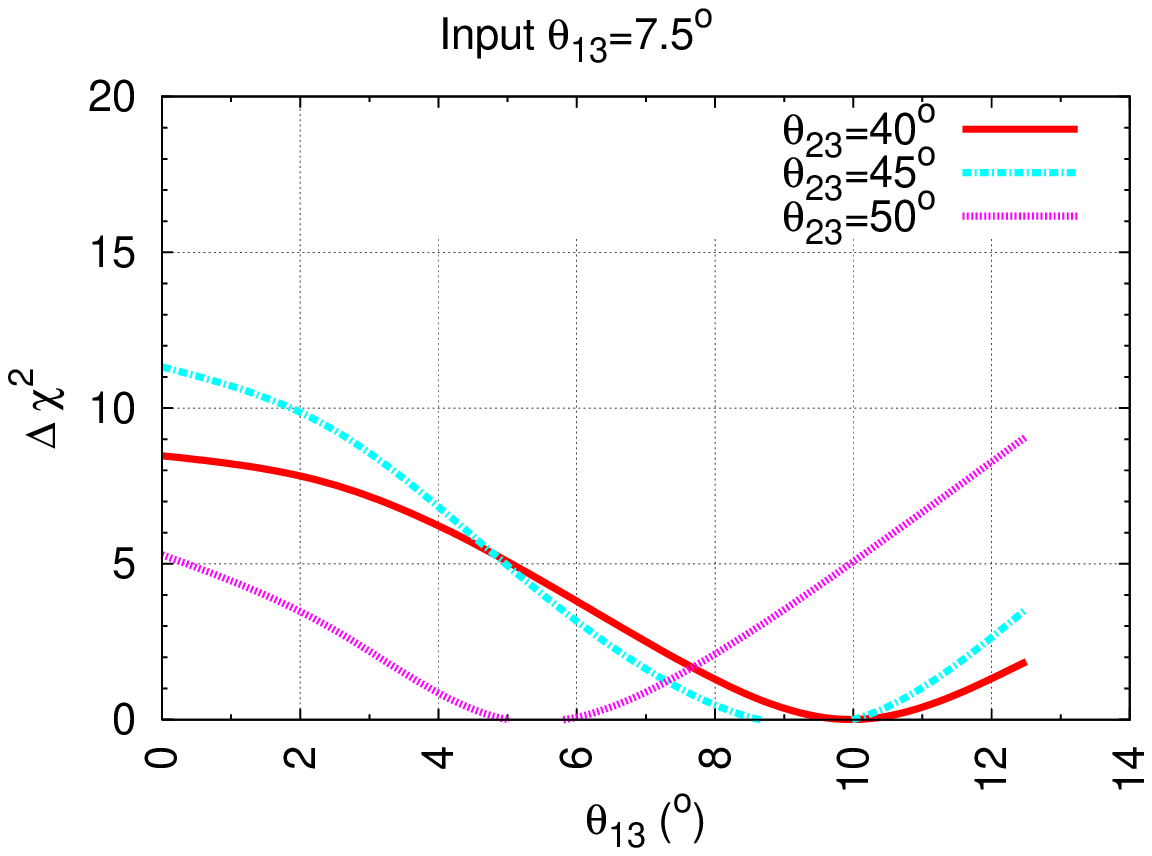}
\caption{\sf \small 
The same as Fig. \ref{f:tt23}, but with $\theta_{13}$.
}
\label{f:tt13}
\end{figure*}
\begin{figure*}[htb]
\includegraphics[width=6.0cm,angle=0]{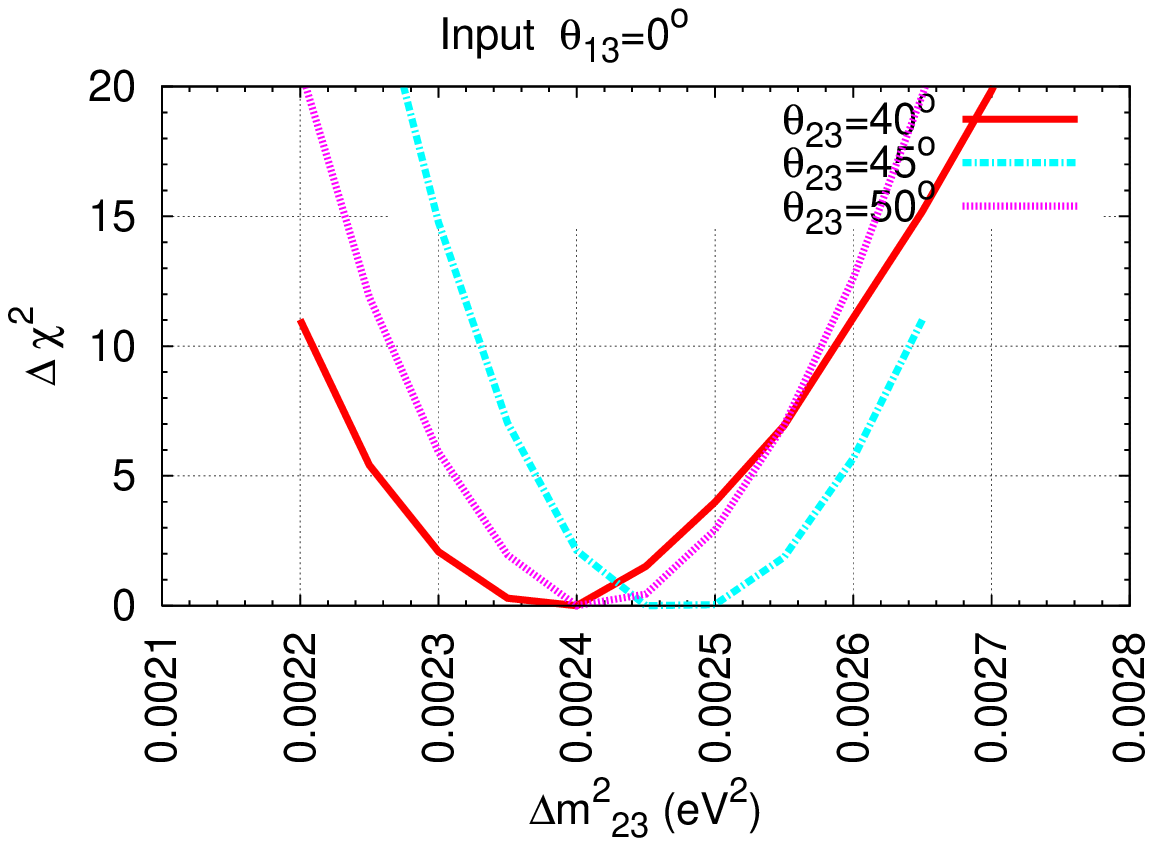}
\includegraphics[width=6.0cm,angle=0]{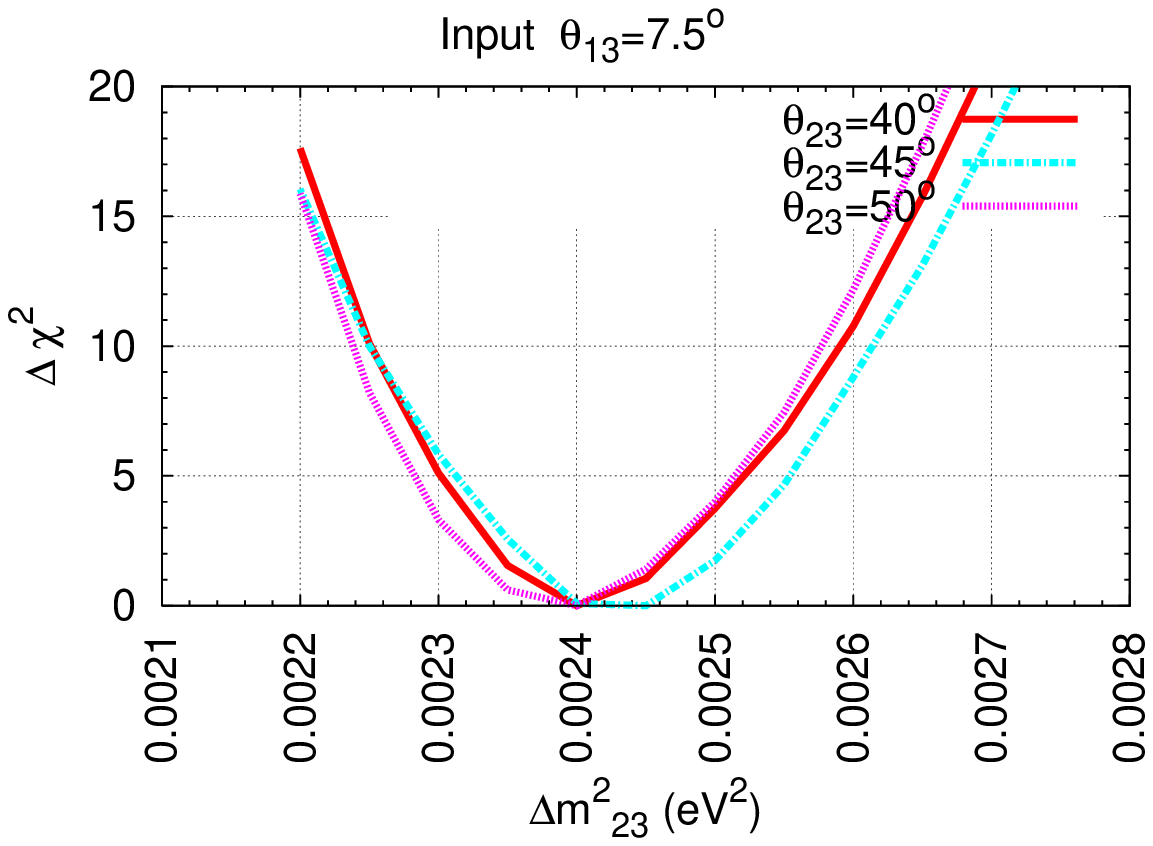}
\caption{\sf \small 
The same as Fig. \ref{f:tt23}, but with $\Delta m^2_{32}$.
}
\label{f:m32}
\end{figure*}

The variation of $\Delta\chi^2[=\chi^2-\chi^2_{\rm min}]$ with each of $\theta_{23},~\theta_{13}$ and $\Delta m^2_{32}$,
are shown in Fig. \ref{f:tt23}, \ref{f:tt13ns}, \ref{f:tt13}, and \ref{f:m32}. 
These are after marginalization over all the oscillation parameters except one, with
which it varies. We present the cases for inputs $\theta_{13}=0^\circ$,and $7.5^\circ$
with $\theta_{23}=40^\circ~,45^\circ$, and $50^\circ,$ respectively. 


\subsection{Sensitivity to $\theta_{23}$ and its octant discrimination}
As the present experiments indicate that the value of $\theta_{13}$ is small 
compared to $\theta_{23}$, the atmospheric neutrino oscillation is mainly 
governed by two flavor oscillation
$\nu_\mu~(\bar\nu_\mu)~\leftrightarrow\nu_\tau~(\bar\nu_\tau)$.
This  constrains $\sin^2 2\theta_{23}$ and $|\Delta m_{32}^2|$. 

\begin{figure*}[htb]
\bc
\vspace*{-0cm}
\includegraphics[width=17cm,height=6.5cm,angle=0]{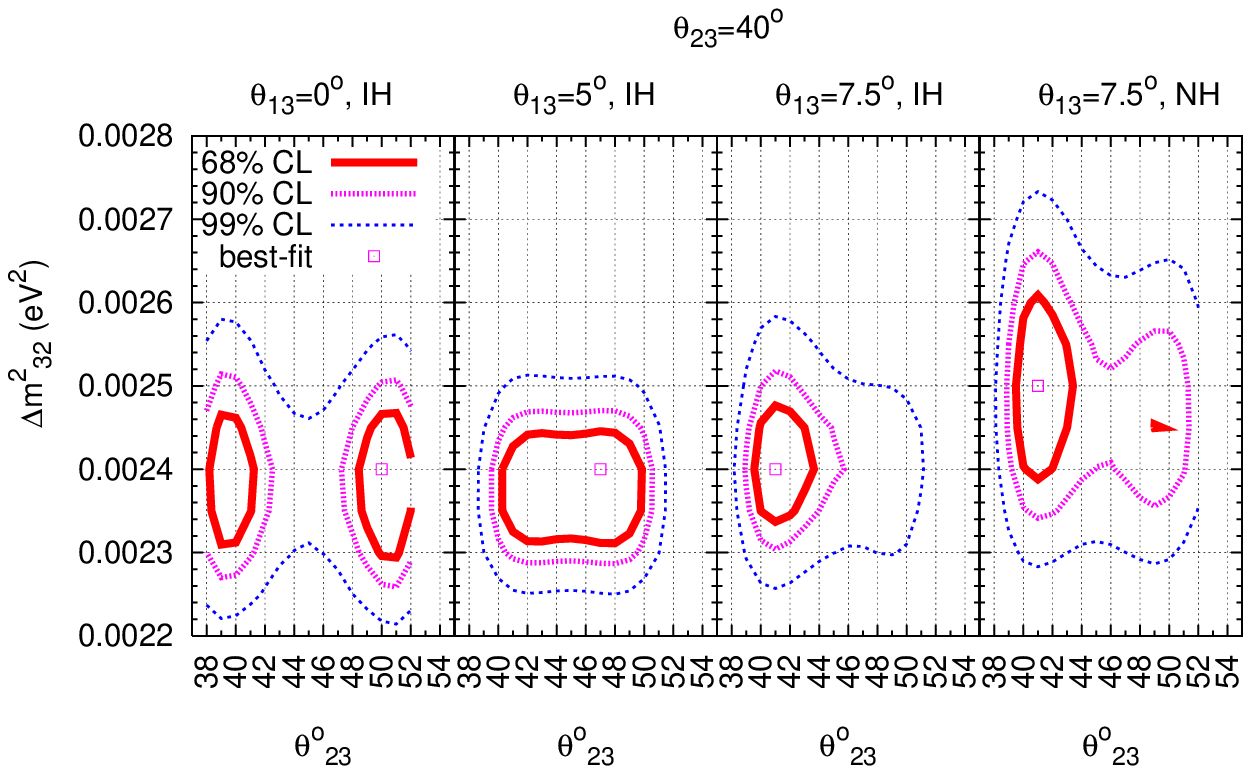}
\vspace*{-0cm}
\includegraphics[width=17cm,height=6.5cm,angle=0]{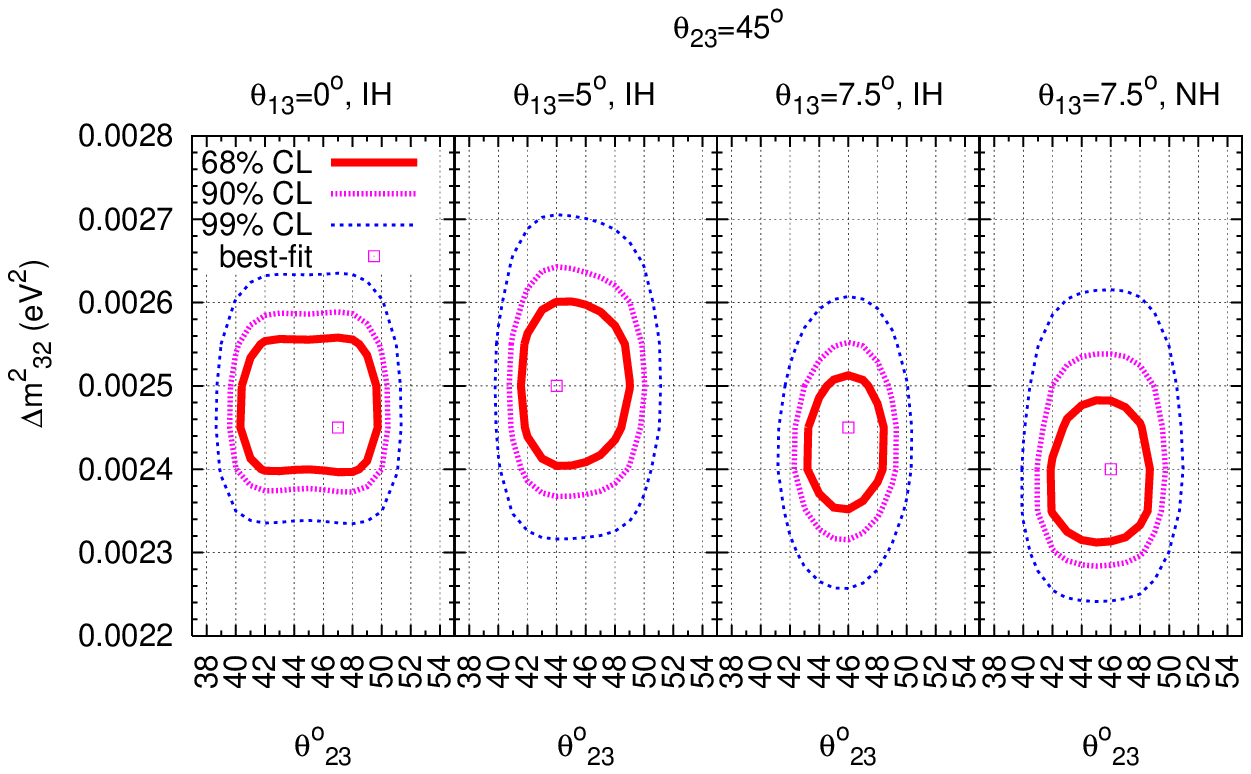}
\vspace*{-0cm}
\includegraphics[width=17cm,height=6.5cm,angle=0]{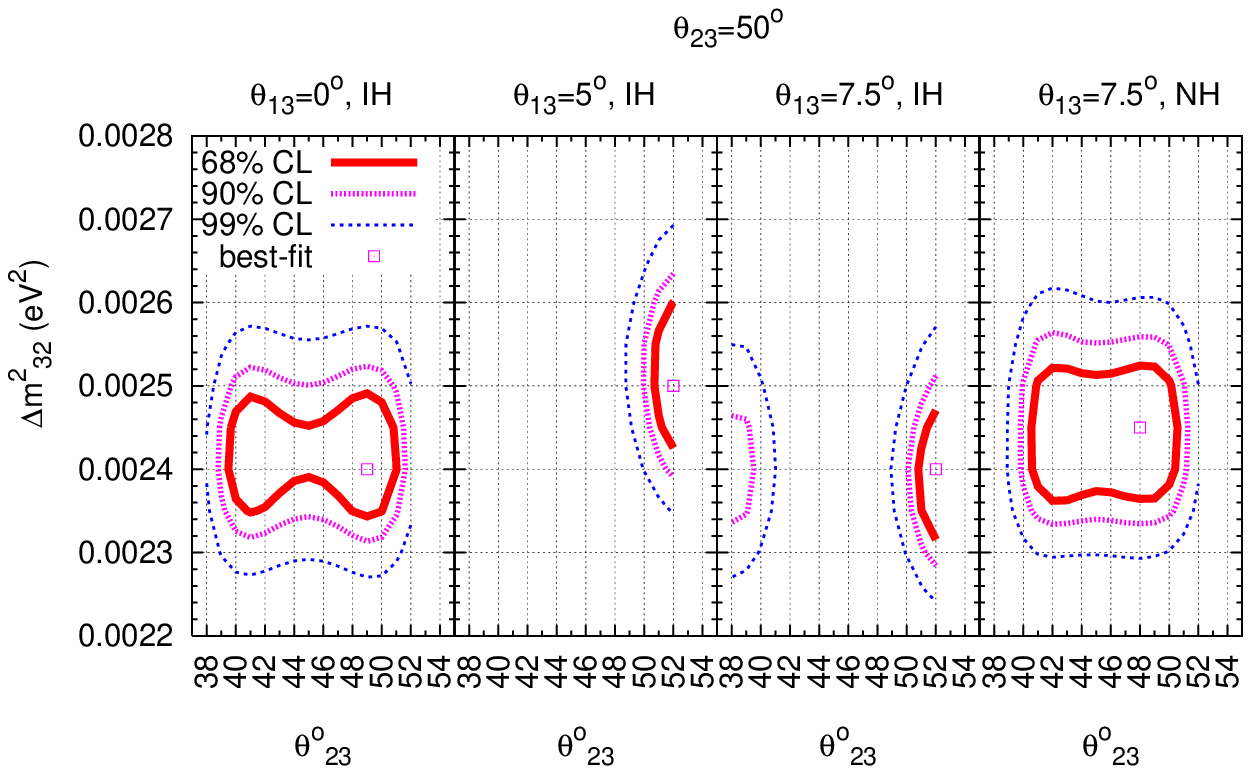}
\ec
\caption{\sf \small 
The 68\%, 90\%, 99\% CL allowed regions in 
$\theta_{23}-|\Delta m_{32}^2|$ plane for the input of  $\theta_{23}=40^\circ$ (first row),
$45^\circ$ (second row), $50^\circ$ (third row) with
$\theta_{13}=0^\circ$ (first column), $5^\circ$ (second column), $7.5^\circ$
(third column) with IH  and $7.5^\circ$ (fourth 
column) with NH. 
}
\label{f:t23m}
\end{figure*}

From Fig. \ref{f:t23m}, 
we see that the deviation from the maximal mixing between {\bf 2} and {\bf 3} flavor eigen 
states  can be observed. 
However, a degeneracy in $\theta_{23}$ arises in case of $\theta_{13}=0$, 
whether it is larger or smaller than $45^\circ$. But, when the matter effect 
comes into the play, a  resonance occurs in 
$\nu_\mu~(\bar\nu_\mu)~\leftrightarrow\nu_e~(\bar\nu_e)$ oscillation and it leads to a large
effective value of $\theta_{13}$  (see Eq. \ref{e:dm31}). 
This helps to dominate the $\sin^4\theta_{23}$ term in Eq. \ref{e:pmumu} and breaks 
the $\theta_{23}$ degeneracy in its measurement.  
Since the atmospheric neutrinos cover a large region of $E-L$ plane, it can 
observe the matter resonance and has an ability to discriminate the octant degeneracy.
In Fig. \ref{f:tt23}, the variations  of $\Delta \chi^2 = (\chi^2-\chi^2_{\rm min})$ with $\theta_{23}$
are shown for input values of $\theta_{23}=40^\circ,~45^\circ$ and $50^\circ$ with $\theta_{13}=0^\circ$
and $7.5^\circ$, respectively. We see that with increase in $\theta_{13}$, the matter effect
not only discriminates the octant, but  increases the precision also. 

In Fig. \ref{f:t23m} we see that for $\theta_{13} =7.5^\circ$ the octant discrimination is 
possible for input of $\theta_{23}=40^\circ$ and $50^\circ$ with IH. But it is not possible for NH.
Normally, the flux of $\nu_\mu$ is higher than $\bar\nu_\mu$. In case of IH (NH),
$\bar\nu_\mu$ ($\nu_\mu$) is suppressed. The statistics remains high for IH compared 
to NH, which leads better octant discrimination possibility for IH.

\subsection{Sensitivity to $\theta_{13}$} 
The effect of $\theta_{13}$ in oscillation probability
does not appear dominantly neither in atmospheric  nor in solar neutrino oscillation, but as 
a subleading in both oscillations. In case of atmospheric neutrino, its effect is seen 
at a) $E\sim$ 1 GeV for propagation of neutrinos through vacuum as well as through matter (no matter
resonance),    and  b) $E\approx 2-10$ GeV for propagation only through matter (matter resonance). 
The matter effect enhances the difference in oscillation probabilities 
between two $\theta_{13}$ values for neutrinos with NH and for anti-neutrinos
with IH (see Eq. \ref{e:dm31}). 
In Fig. \ref{f:tt13ns} we show the cases a) and b) considering neutrinos and anti-neutrinos separately.
We find that the effect of case a) is negligible.

We have plotted the 
APS in $\theta_{13}-|\Delta m_{32}^2|$ plane 
in Fig \ref{f:t13m} for $\theta_{13}=0^\circ, ~5^\circ,$  and $7.5^\circ$ with $\theta_{23}=40^\circ$,
$45^\circ$ and $50^\circ$, respectively. 
We find that the matter effect significantly constrains $\theta_{13}$ over
the present limit. Though the matter effect acts either on neutrinos
or on anti-neutrinos depending on the type of the hierarchy, but we have checked that it improves when 
we consider both neutrinos and anti-neutrinos. The sensitivity of $\theta_{13}$ is not generally 
expected to be improved for the case of analysis with neutrinos and 
anti-neutrinos in together.  However, this happens here due to  the marginalization 
which  restricts $\theta_{23}$ more tightly
for the case of $\nu$ and $\bar\nu$ in together than either with $\nu$ or $\bar\nu$
and  indirectly constrains $\theta_{13}$. It is also seen that the APS is 
strongly dependent on the input of $\theta_{23}$ and a better constraint is obtained
for $\theta_{23} > 45^\circ$.
However, it is notable here that the uncertainty is very high and the best-fit values
deviate largely from its input values for nonzero $\theta_{13}$ inputs 
due to the reasons discussed at the beginning of this section.


\subsection{Sensitivity to $\Delta m^2_{32}$} 
We show the constraint on $|\Delta m^2_{32}|$ in Fig. \ref{f:t23m} and \ref{f:t13m}.
We see that the precision is little better when $\theta_{13}=0.$ The reason behind this is that
a regular oscillation pattern with periodic rise and fall is observed when $\theta_{13}=0$.



It is  seen that the APS is larger for NH than IH. The matter effect does not act on 
neutrino for IH and anti-neutrinos for NH with an addition to the fact that the flux is higher
for neutrino than anti-neutrino. 
As discussed above, the APS is more restricted when there is no matter effect.
Here, for input with IH 
the number of neutrino events is high and they do not have any  matter effect. 
This leads to smaller APS for IH compared to NH for large values of $\theta_{13}$.
 

\subsection{Effect of events at near horizon on precision measurements}
\begin{figure*}[htb]
\bc
\includegraphics[width=17cm,height=6.cm,angle=0]{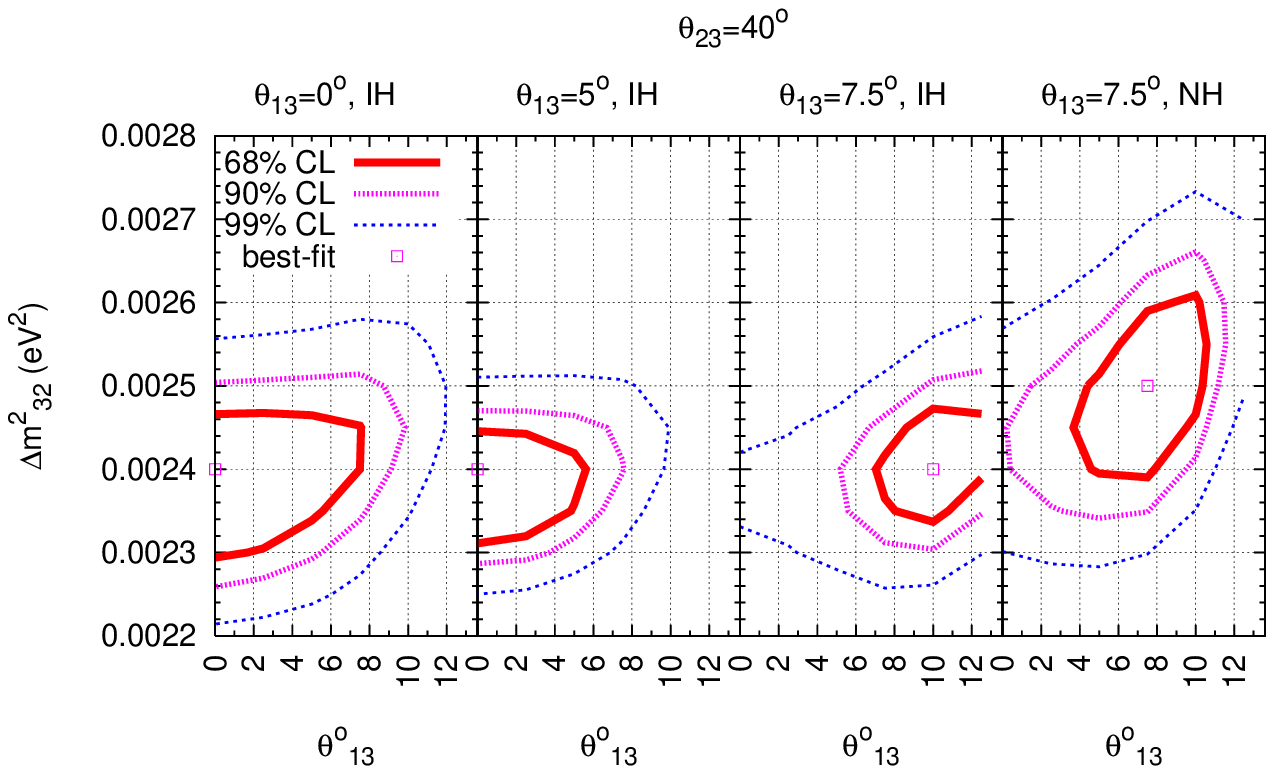}
\includegraphics[width=17cm,height=6.cm,angle=0]{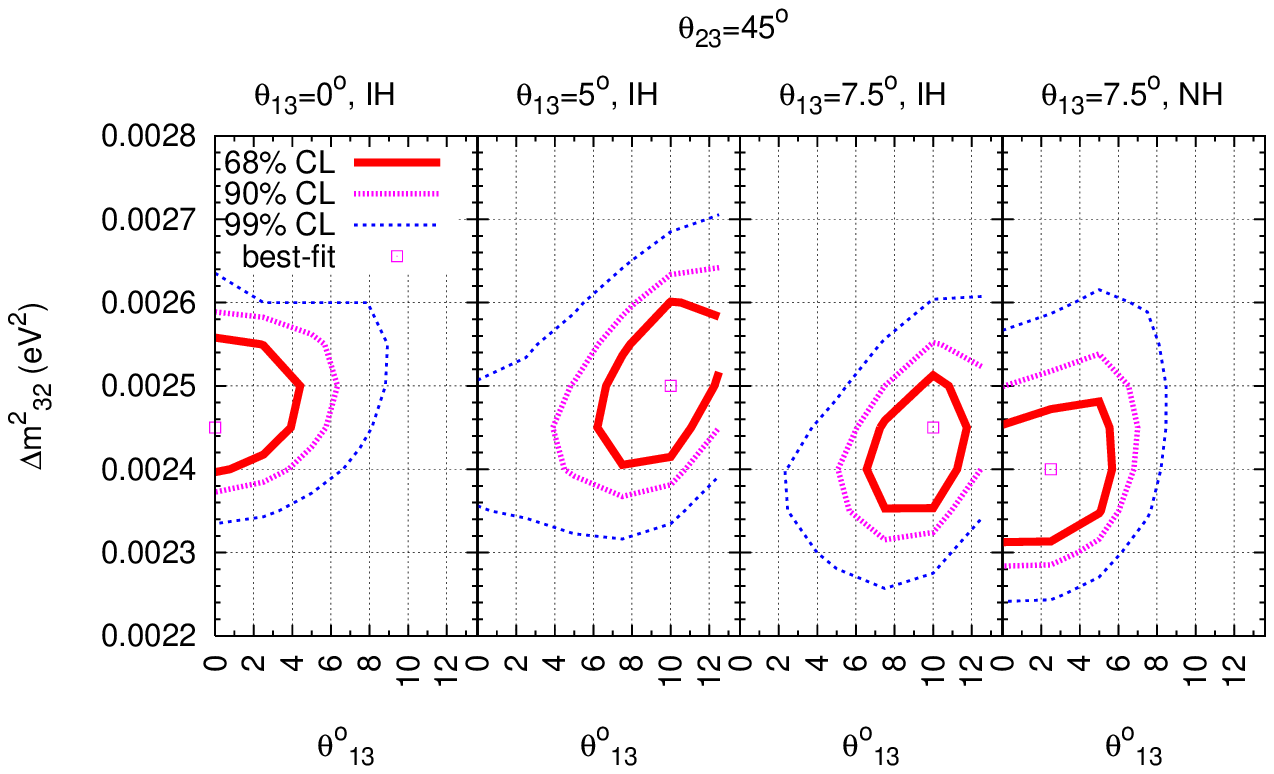}
\includegraphics[width=17cm,height=6.cm,angle=0]{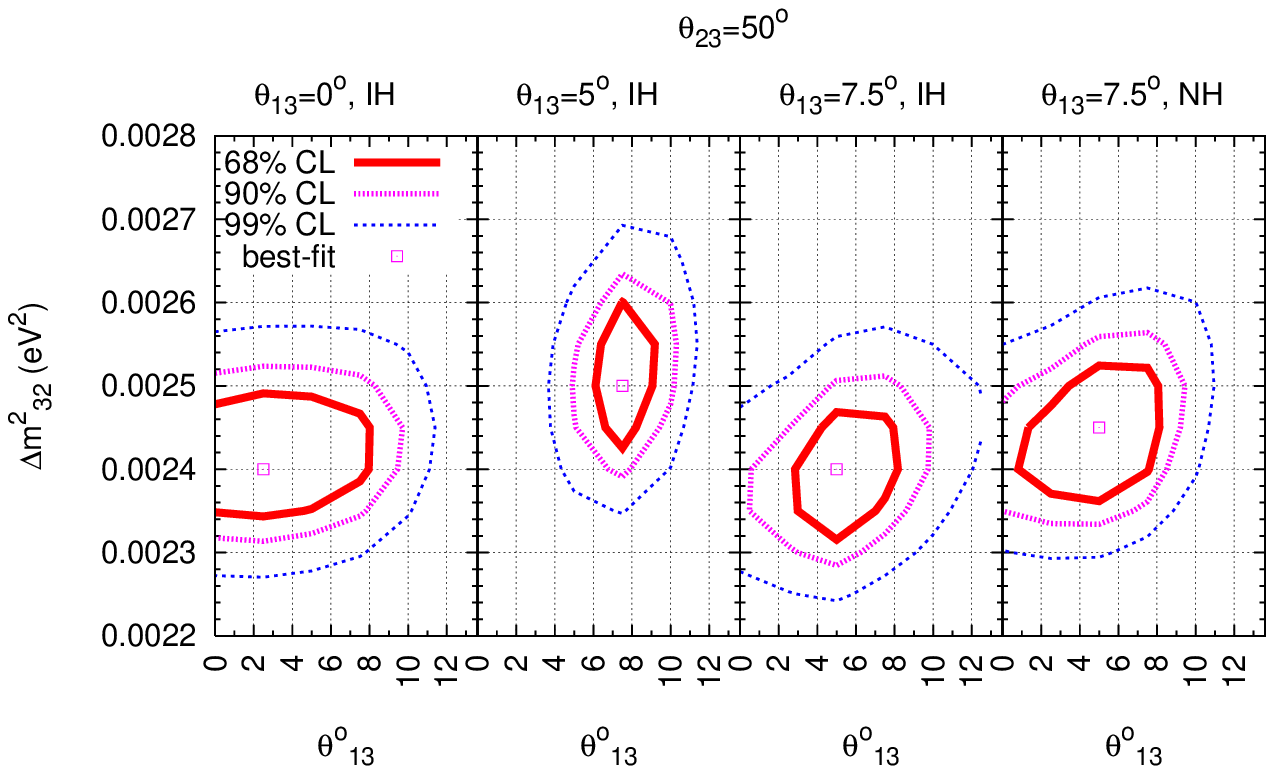}
\ec
\caption{\sf\small 
The same as Fig. \ref{f:t23m}, but in $\theta_{13}-|\Delta m_{32}^2|$ plane.
}
\label{f:t13m}
\end{figure*}

\begin{figure*}[htb]
\bc
\includegraphics[width=8cm,height=8.0cm,angle=0]{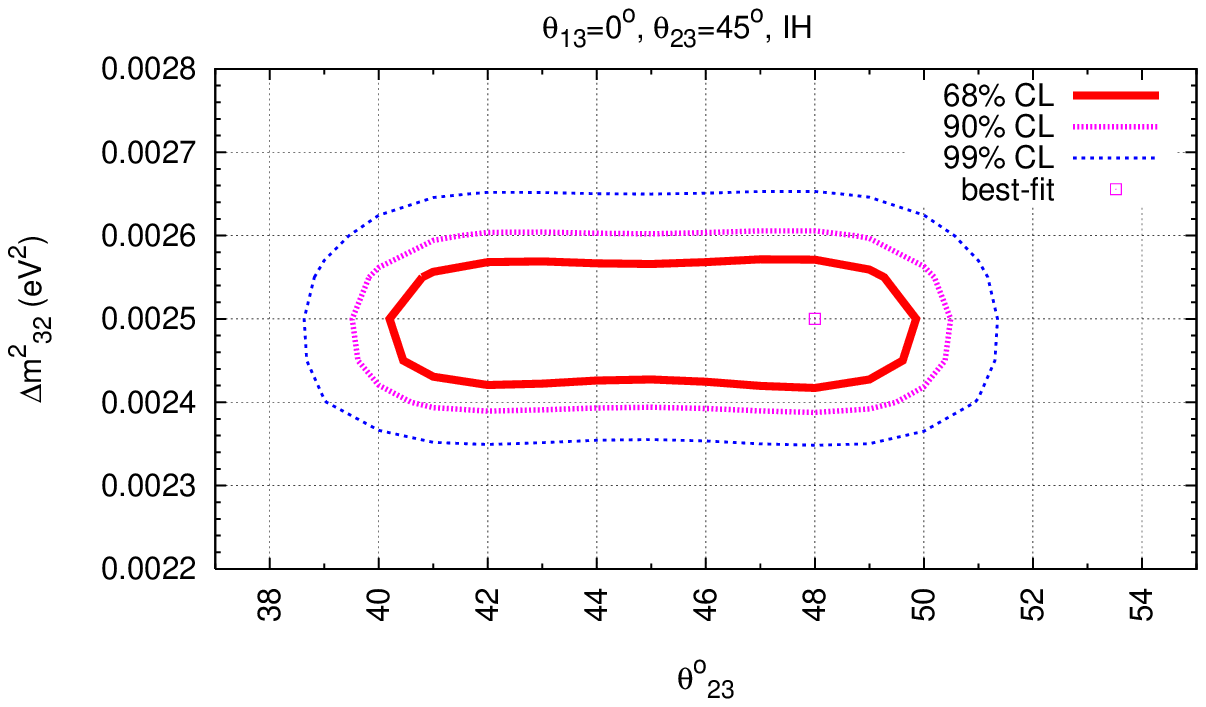}
\includegraphics[width=8cm,height=8.0cm,angle=0]{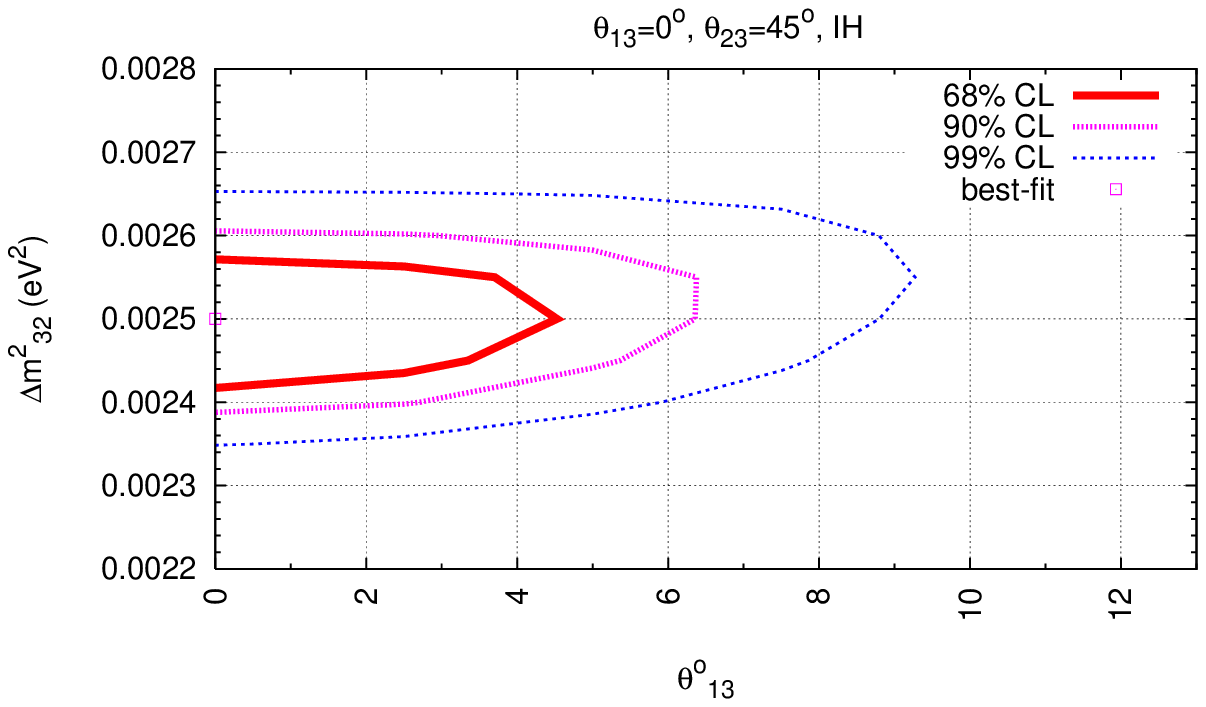}
\ec
\caption{\sf\small
 The  allowed regions without any zenith angle cut for the events at the horizon.
}
\label{f:nocut}
\end{figure*}

For a given set of input parameters, if we compare the APSs with zenith angle cut 
(discussed in section \ref{s:cut}) with those without any cut,
we find no significant differences.
As a demonstrating example, we have shown the APSs in Fig. \ref{f:nocut} without imposing 
any zenith angle cut for a given set of oscillation parameters.   
One can find the corresponding  plots with  zenith angle cut in Figs. \ref{f:t23m} and \ref{f:t13m}. 

From the study of this paper, we can conclude that the events at near horizon cannot
contribute significantly in precision measurements. The fact is that
the $L$ resolution is very poor here. A little change in zenith angle at near horizon changes
$L$ values drastically. Again, the discrimination of up and down going
events are not possible. So, the oscillation effect is almost smeared out by the resolutions.
From the $L/E$ dip considering the $L$ and $E$
values of neutrinos, one can expect a large contribution in precision
from these events. But, in practical situation, there is no appreciable improvement
after addition of these events.

The vertical (horizontal) stacking of iron plates will be able to detect the horizontal
(vertical)  events. So, from this study one can conclude that horizontal stacking is 
expected to give better precision than the vertical stacking.

\subsection{Precision of the parameters}
For a quantitative assessment of the result, we define the precision 
of a parameter $t$ as:  
\be P=2\left(\frac{t^{\rm max}-t^{\rm min}} {t^{\rm max}+t^{\rm min}}\right ).\ee
We find that the precisions are strongly dependent on the set of input parameters.
We obtained the precision of 
$|\Delta m_{32}^2| \approx 6.4\%,~  8.8\%$ and $12\%$ 
at 68\%, 90\% and 99\% CL, respectively and 
the precision of $\sin^2\theta_{23}\approx  31\%$, 38\%, and 
 41\% at 68\%, 90\%
and 99\% CL, respectively for the  input of $\theta_{23}=45^\circ$ and $\theta_{13}=0^\circ$. 

The oscillation dip moves towards the lower $L/E$
values as $|\Delta m_{32}^2|$ increases. The statistics also decreases at 
the lower $L/E$ region. So, the precision is expected  to be weaker  as the input of  $|\Delta m_{32}^2|$ increases.

%
A comparison of the precisions of $\Delta m^2_{32}$ and $\sin^2\theta_{23}$ among different future 
baseline experiments is made in \cite{Huber:2004dv}. The variation of the precisions with 
the change of input parameters are also presented there.  
We compare  our results with 5 years run of T2K, which is the best in determining precision 
of atmospheric oscillation parameters in the list in \cite{Huber:2004dv}. 
The precision of $\Delta m^2_{32}$ is almost same 
with T2K ($\approx 12\%$) and precision of $\sin^2\theta_{23}$ from ICAL is 41\% while from T2K is 46\%.
Here we present the results for 10 years run of 100 kTon ICAL detector.
From this work it is also seen that atmospheric neutrinos
at ICAL detector are in very good position to discriminate octant of $\theta_{23}$.
The main advantage here is that atmospheric neutrinos are natural sources
and the cost goes only to build and run the detector.




\section{Conclusion}
We have studied the precisions of the oscillation parameters from
atmospheric neutrino oscillation experiment at the large magnetized ICAL detector
generating events by Nuance and considering only the muons produced by the charge
current interactions. 
The distance between two consecutive peaks of oscillation
in $E$ for fixed $L$ increases as one goes from higher $L$ values 
to its lower values. This indicates the need of finer binning at lower 
$L$ values in $\chi^2$ analysis. We optimize the binning of the data
in the grids of $\log E - L^{0.4}$ plane.
We find that the impact of the events at near horizon on the precision measurements
is very negligible due to poor $L$ resolution.

From the marginalized $\chi^2$ study 
separately for neutrinos and anti-neutrinos, we find that
the measurement of $\theta_{13}$  is possible at a considerable precision
with atmospheric neutrinos.
The precision of $\theta_{13}$  depends crucially on its input value.
For $\theta_{13}=0$, we find its  upper bound  $\approx 4^\circ,~ 6^\circ$ 
and $9^\circ$ at 68\%, 90\% and  99\% CL, respectively. 
The both lower and upper bounds of $\theta_{13}$ are also possible for some 
combinations of ($\theta_{23}, \theta_{13}$) and it happens mainly 
for $\theta_{23}\gapp 45^\circ$.

The precision of $|\Delta m_{32}^2|$ and $\theta_{23}$ can also be very
high and the determination of octant of $\theta_{23}$ is possible
for some combinations of ($\theta_{23},~\theta_{13}$).

It should also be noted here that in $\chi^2$ analysis the theoretical
data and the experimental data are not generated in the same way.
The  different models of neutrino interactions generate  energy
dependent systematic uncertainties at some energies.
These are not included in this analysis. This causes
sometimes  large deviation of the best-fit values of the oscillation parameters
from the input values.

{\bf Acknowledgments:}
This research has been supported by funds from Neutrino Physics projects 
at HRI. The use of excellent cluster computational facility installed from this project 
 is  gratefully acknowledged. A part of the computation was also carried out in HRI 
general cluster facility.

\end{document}